\documentclass[11pt,a4paper]{article}
\usepackage{amssymb}
\usepackage{amsmath}
\usepackage{graphicx}

\pdfoutput=1

\begin{document}

\begin{center}
\textbf{CONSTRAINT ON DARK MATTER CENTRAL DENSITY IN THE EDDINGTON INSPIRED BORN-INFELD (EiBI) GRAVITY WITH INPUT FROM WEYL GRAVITY}

\bigskip

\bigskip

Alexander A. Potapov$^{1,a}$, Ramil Izmailov$^{2,b}$, Olga Mikolaychuk$^{1,c} $, 

Nikolay Mikolaychuk$^{1,c}$, Mithun Ghosh$^{3,d}$ and Kamal K. Nandi$^{1,2,3,e}$

$\bigskip $

$^{1}$Department of Physics \& Astronomy, Bashkir State University,
Sterlitamak Campus, Sterlitamak 453103, RB, Russia

$^{2}$Zel'dovich International Center for Astrophysics, M. Akmullah Bashkir
State Pedagogical University, Ufa 450000, RB, Russia \\[0pt]
$^{3}$ Department of Mathematics, University of North Bengal, Siliguri
734013, WB, India \\[0pt]

\bigskip

$\bigskip $

$^{a}$Email: potapovaa@mail.ru

$^{b}$Email: izmailov.ramil@gmail.com

$^{c}$Email: mikov94@mail.ru

$^{d}$Email: ghoshmithun123@gmail.com

$^{e}$Email:kamalnandi1952@yahoo.co.in

\bigskip
\end{center}

PACS\ number(s): 04.50.Kd, 95.30.Sf, 04.50.1h

\begin{center}
\textbf{Abstract}
\end{center}

Recently, Harko \textit{et al.} (2014) derived an approximate metric of the
galactic halo in the Eddington inspired Born-Infeld (EiBI) gravity. In this
metric, we show that there is an upper limit $\rho _{0}^{\text{upper}}$ on
the central density $\rho _{0}$ of dark matter such that stable circular
orbits are possible \textit{only} when the constraint $\rho _{0}\leq \rho
_{0}^{\text{upper}}$ is satisfied in each galactic sample. To quantify
different $\rho _{0}^{\text{upper}}$ for different samples, we follow the
novel approach of Edery \& Paranjape (1998), where we use as input the
geometric halo radius $R_{\text{WR}}$ from Weyl gravity and equate it with
the dark matter radius $R_{\text{DM}}$ from EiBI gravity for the same halo
boundary. This input then shows that the known fitted values of $\rho _{0}$
obey the constraint $\rho _{0}\leq \rho _{0}^{\text{upper}}\propto $ ($R_{%
\text{WR}}$)$^{-2}$. Using the mass-to-light ratios giving $\alpha $, we
shall also evaluate $\rho _{0}^{\text{lower}}$ $\propto $ $(\alpha -1)M_{%
\text{lum}}R_{\text{WR}}^{-3}$ and the average dark matter density\ $%
\left\langle \rho \right\rangle ^{\text{lower}}$. Quantitatively, it turns
out that the interval $\rho _{0}^{\text{lower}}$ $\leq \rho _{0}\leq $ $\rho
_{0}^{\text{upper}}$ verifies reasonably well against many dark matter
dominated low surface brightness (LSB) galaxies for which values of $\rho
_{0}$ are independently known. The interval holds also in the case of Milky
Way galaxy. Qualitatively, the existence of a stability induced upper limit $%
\rho _{0}^{\text{upper}}$ is a remarkable prediction of the EiBI theory.

\bigskip


\begin{center}
\bigskip

\textbf{I. INTRODUCTION}
\end{center}

Early observations [1-3] on rotational data of spiral galaxies, now
reconfirmed by observations extending well beyond the optical disc [4-18],
indicate that they do not conform to Newtonian gravity predictions. Hence
the hypothesis is that there could be large amounts of non-luminous matter
hidden in the galactic haloes. The rationale is this: Doppler emissions from
stable circular orbits of neutral hydrogen clouds in the halo allow
measurement of tangential velocity $v_{\text{tg}}$ of the clouds treated as
probe particles. Contrary to Newton's laws, where $v_{\text{tg}}^{2}$ should
decay with radius $r$, observations indicate that it approximately levels
off with $r$ \ in the galactic halo region, which in turn calls for the
presence of additional non-luminous mass, the so called dark matter. Since
dark matter has not yet been directly observed, the dark matter hypothesis
is often variantly referred to as the missing mass problem.

Several well known theoretical models for dark matter exist in the
literature but it is impossible to list all of them here (only some are
mentioned in [19-34]). In its usual formulation, dark matter is a
parametrization of the observed velocity discrepancies and is not a
prediction of the formulation. Some simulations require fine tuning of halo
parameters to luminous parameters galaxy by\ galaxy $-$ a procedure that
only enlarges the number of parameters rather than reducing them (See [35],
pp.32-33; see also footnote 13). There exist yet another variety of halo
models, which treat the missing mass problem as a failure of the Newtonian
theory on galactic distance scales rather than as a prediction for dark
matter. Such models actually do \textit{not} require dark matter at all for
the interpretation of observed rotation data. This class of theories
include, e.g., Modified Newtonian Dynamics (MOND) developed by Milgrom [36-42],
Scalar-Tensor-Vector Gravity theory developed by Moffat [43-45], Weyl conformal
gravity\footnote{%
Weyl conformal gravity has been debated for and against in the literature.
For instance, Flanagan [46] argues that if the source has associated with it
a macroscopic long range scalar field, breaking conformal symmetry, the
theory does not reproduce attractive gravity in the solar system. However,
subsequently, Mannheim [47] has counter-argued that Schwarzschild tests of
solar gravity could still be recovered even in the presence of such
macroscopic fields.} implemented by Mannheim and O'Brien [48]. For brevity,
we call the last the MO model and we shall use this ingredient in the
sequel. A remarkable speciality of the MO model is that, using the best
available galactic optical and radio data, and a standardized, non-biased,
treatment for selecting appropriate galactic parameters, the model is able
to provide a good fit to the rotation curves \textit{without} the need for
any dark matter whatsoever.

There are various other (non-)dark matter models\ that are capable of
accounting, for example, for observations of galaxy clusters and
gravitational lensing or structure formation. \ A leading example is the
cold dark matter (CDM)\ model, which is a part of the current standard model 
$\Lambda $CDM of cosmology. These models are based on different
phenomenologies such as inflation and nucleosynthesis [49-66]. They can
successfully explain observations of galaxy clusters [49-54], gravitational
lensing [55,56] or structure formation [57], to name the most important ones.
These models postulate that galactic cores may consist of axions [58],
massive gravitons [59], BEC [60] or other collisionless particles. The
post-recombination fluctuation spectrum nicely explains the formation of
galaxies and clusters [49-54]. The CDM is a successful paradigm accounting for
the small density inhomogeneities that seed structure formations $10^{-34}$
sec after the bang and as such provides a bold probe into the Early Universe
[57]. Some\ other and recent works on CDM models are mentioned here though
the list by no means is exhaustive [61-66].

Recently, another alternative candidate for dark matter is also being
speculated. This is based on the evidence of soft positron spectrum in the
AMS-02 [67,68] cosmic ray data. Despite this alternative, the observed flat
rotation curves are still considered as a robust proof that dark matter
essentially is of gravitational origin described by general relativity [69-71].
But in general relativity, matter-gravity coupling is linear, while some
authors argue (for non-minimal coupling of modified gravity with matter or
other insights into the paradigm, see [72-77]) that there is no obvious reason
as to why the coupling should be linear. Following this thought, an
interesting modification of matter-gravity coupling leading to the
Eddington-inspired Born-Infeld (EiBI) theory has been recently developed by
Ba\~{n}ados and Ferreira [78,79], which we shall use as another ingredient.
Only in vacuum, the EiBI theory is equivalent to standard general
relativity. This new, and more general, theory has led to interesting
observable predictions in the context of solar interior dynamics, big bang
nucleosynthesis, neutron stars, the structure of other compact stars [80-83]
including the possibility of nonsingular cosmological models and alternative
to inflation [84].

To get a more complete view of the two ingredients of the present work, the
EiBI and the Weyl gravity, it is necessary to mention that several important
predictions follow from the two theories. For instance, recent investigation
by Du \textit{et al. }[85] on large scale structure formation in the EiBI
gravity shows a deviation in the matter power spectrum between the EiBI
gravity and the $\Lambda $CDM~model, which is a testable prediction.
Stability and localization of gravitational fluctuations in the EiBI brane
system have been studied in Refs. [86,87]. Further, as shown by Wei \textit{%
et al. }[88], strong gravitational lensing observables in EiBI are
controlled by the coupling parameter $\kappa $, which is a new prediction
that lends itself to future testing. Similarly, in Weyl gravity, the first
order light deflection angle $\theta _{\text{W}}$ by a galaxy, first
obtained by Edery and Paranjape [89], contains the galactic halo parameter $%
\gamma $ appearing in the MO model. Bhattacharya \textit{et al.} [90,91]
calculated higher order deflection terms. Strong field lensing in the Weyl
gravity has been studied recently in [92], and the predictions can be
verified by actual observations in future. A remarkable feature of Weyl
gravity is that its solution, the MO model, already contains the successes
of the well tested Schwarzschild gravity as a special case. All the above
exemplify the current status of the capabilities of the two theories in
question.

The possibility of perfect fluid dark matter within the framework of general
relativity has already been explored in the literature [93,94]. A similar
possibility has been recently investigated within the framework of the EiBI
theory by Harko \textit{et al.} [95,96] and this is the model we are going to
analyze further in this paper. Using a\ tangential velocity profile [97,98]
giving Universal Rotation Curves (URC) and setting the cosmological constant
to zero, they obtained, in the Newtonian approximation, a new galactic
metric and theoretically explored its gravitational properties. However, the
numerical values of the crucial parameter $\kappa $ (denoted by $\kappa =2R_{%
\text{DM}}^{2}/\pi ^{2}$) or equivalently the dark matter radius $R_{\text{DM%
}}$, cannot be determined from the theory alone $-$ it has to be obtained
either from the observed data or from some other model.\footnote{%
We wish to clarify that the EiBI parameter $\kappa $ is not a universal
constant $-$ it's more like a parameter of the theory that assumes different
values depending on the physical situation. For instance, the structure of
compact general relativistic star requires a value $\kappa \simeq 10^{12}$ cm%
$^{2}$ [82], which differs from the value $\kappa \simeq 10^{44}$ cm$^{2}$
inferred from dark matter density profiles.
\par
{}} It is also expected that the values of $R_{\text{DM}}$ would differ from
galaxy to galaxy. On the other hand, to our knowledge, apart from the
observed last scattering radii $R_{\text{last}}$, the astrophysical
literature still seems to lack concrete observed data on $R_{\text{DM}}$ for
individual galaxies. Therefore, an appropriate numerical input for $R_{\text{%
DM}}$ is needed, which we take from Weyl gravity, if we want to make \textit{%
quantitative} predictions.

At this point, we recall a novel idea of Edery \& Paranjape [89], where they
bridged two different metric theories by equating the \textit{same} Einstein
angle $\theta _{\text{E}}$ (caused by the luminous + dark matter) with the
Weyl angle $\theta _{\text{W}}$ (caused by the luminous matter alone), and
drew useful and testable conclusions using the identity $\theta _{\text{E}%
}=\theta _{\text{W}}$. Motivated exactly by this idea, we equate the \textit{%
same} EiBI radius of dark matter $R_{\text{DM}}$ (caused by dark matter
source) with the geometric Weyl radius of the galactic halo $R_{\text{WR}}$
(caused by the luminous matter alone). With the numerical input $R_{\text{DM}%
}=R_{\text{WR}}$, we shall quantify the relevant central densities in the
EiBI theory (see footnote 8). \textit{We wish to clarify that we are not
talking here of merging or mapping the two theories into one another per se }%
but concentrating only on a particular common prediction. The theories are
of course different from each other $-$ one with dark matter source and the
other without, not to mention differences elsewhere. But both are metric
theories capable of predicting for \textit{any} given galactic sample\textit{%
\ }a dark matter/halo boundary arising out of the same stability condition $%
V^{\prime \prime }<0$ (as used, e.g., in the braneworld dark matter [99]).
Therefore, without any bias to either theories, we shall investigate if this
input leads to limits on dark matter central density $\rho _{0}$ consistent
with those estimated from fits to different known profiles. We shall see
that it does.

The radius $R_{\text{WR}}$ is to be understood as the geometric halo radius
with its interior being filled with Weyl vacuum.\footnote{%
Note that we are using here the terminology $R_{\text{WR}}$ in lieu of $R_{%
\text{stable}}^{\text{max}}$ of Ref.[100] only to bring it in line with the
notation of the present analysis.} We stress that Weyl vacuum is \textit{not}
a vacuum in the ordinary sense but an arena of interplay of several
potential energies, predominantly the global quadratic potential due to
cosmic inhomogeneities [48]. Thus, our input physically means that the total
potential energy contained within the halo radius $R_{\text{WR}}$ of Weyl
gravity equals the total invisible dark matter energy contained within $R_{%
\text{DM}}$ of EiBI gravity. The radius $R_{\text{DM}}$ is defined by the
absence of dark matter density at the halo boundary [95,96], while $R_{\text{WR}%
}$ is defined by the absence of stable circular orbits at the halo boundary
[100]. Since stability is an essential physical criterion because Doppler
emissions from the halo emanate from stable circular orbits of hydrogen gas
[101], we think that $R_{\text{WR}}$ should be regarded as the only testable
upper limit on the radius of a galactic halo. Fortunately, observed last
scattering data $R_{\text{last}}$ so far have not surpassed the predicted
limiting value $R_{\text{WR}}$ for all the galaxies studied to date, thereby
lending excellent observational support to the MO model prediction of $R_{%
\text{WR}}$.

The purpose of the present paper is as follows: We shall concentrate on the
low surface brightness (LSB) galaxies that are mostly dominated by dark
matter. We show that there is an upper limit on the dark matter central
density $\rho _{0}$ specific to each individual galaxy, which we call here $%
\rho _{0}^{\text{upper}}$, such that stable circular orbits in the EiBI are
possible \textit{only} when the constraint $\rho _{0}\leq \rho _{0}^{\text{%
upper}}\propto $ ($R_{\text{WR}}$)$^{-2}$ is satisfied in that galaxy. Using
the Weyl gravity input $R_{\text{DM}}=R_{\text{WR}}$, we shall then quantify 
$\rho _{0}^{\text{upper}}$, and show that the central density $\rho _{0}$
predicted from the fit to various simulations including the NFW and Burkert
density profiles does obey the constraint. Taking into account the range of
the rather uncertain but possible mass-to-light ratios (denoted, say, by $%
\alpha $), we shall calculate also the lower central density $\rho _{0}^{%
\text{lower}}\propto (\alpha -1)M_{\text{lum}}R_{\text{WR}}^{-3}$ and find
that $\rho _{0}^{\text{lower}}\leq \rho _{0}\leq \rho _{0}^{\text{upper}}$
holds for some known LSB samples for which $\rho _{0}$ is known from
independent fits. Some illustrative galactic samples including the Milky Way
are tabulated. The values fall within the predicted interval for each
individual galaxy and we conjecture that at least the upper limit might be
generally true.

The contents are organized as follows: Since both the models under
consideration are relatively new, hence possibly unfamiliar, we provide in
Sec.II, a brief outline of the algorithms underlying the EiBI and MO metric
models of the galactic halo. In Sec.III, we shall graphically explore $\rho
_{0}^{\text{upper}}$ for stability of circular orbits in EiBI. In Sec.IV,
using the input under consideration, we shall quantify upper and lower
central densities for some samples to see if the estimated central densities 
$\rho _{0}$ truly fall within the proposed interval, that is, if $\rho _{0}^{%
\text{lower}}\leq \rho _{0}\leq \rho _{0}^{\text{upper}}$. In Sec.V, we
discuss the dark matter density profiles in the context of Milky Way. The
results are summarized in Sec.VI. We shall take units such that $8\pi G=1$, $%
c=1$, unless restored.

\begin{center}
\textbf{II (a): EiBI MODEL}
\end{center}

For easy reference, we outline only the salient features of the EiBI dark
matter model developed in [95,96]. The EiBI action is [78,79,85,102,103] 
\begin{equation}
S_{\text{EiBI}}=\frac{2}{\kappa }\int d^{4}x\left[ \sqrt{-\left\vert g_{\mu
\nu }+\kappa R_{\mu \nu }(\Gamma )\right\vert }-\lambda \sqrt{-\left\vert
g_{\mu \nu }\right\vert }\right] +S_{\text{matter}}\text{,}
\end{equation}%
where $\lambda $ is a dimensionless parameter, $g_{\mu \nu }$ is the
physical metric, $R_{\mu \nu }(\Gamma )$ is the symmetric part of the Ricci
tensor built solely from the connection $\Gamma _{\beta \gamma }^{\alpha }$ $%
\left( \equiv \frac{1}{2}q^{\alpha \sigma }\left[ \partial _{\gamma
}q_{\sigma \beta }\text{ }+\partial _{\beta }q_{\sigma \gamma }-\partial
_{\sigma }q_{\beta \gamma }\right] \right) $ derived from an auxiliary
metric denoted by $q_{\mu \nu }$. The meaning of the auxiliary metric $%
q_{\mu \nu }$ is that it partially satisfies Eddington field equations so
that $2\kappa \sqrt{\left\vert R\right\vert }R^{\mu \nu }=\sqrt{\left\vert
q\right\vert }q^{\mu \nu }$, which can be rewritten as Einstein field
equations if we equate $q_{\mu \nu }$ with $g_{\mu \nu }$ and $\kappa $ with 
$\Lambda ^{-1}$. For small values of $\kappa R$, the action (1) reproduces
the Einstein-Hilbert action with $\lambda =\kappa \Lambda +1$, where $%
\Lambda $ is the cosmological constant, while for large values of $\kappa R$%
, the action approximates to that of Eddington, viz., $S_{\text{Edd}%
}=2\kappa \int d^{4}x\sqrt{\left\vert R\right\vert }$. For more details, see
[78,79]. 

Harko \textit{et al.} [95,96] deals with dark matter modeling assuming certain
restrictive conditions such as spherical symmetry and asymptotic flaness,
the latter requiring that $\Lambda =0\Leftrightarrow \lambda =1$. These
assumptions of course limit the applicability of EiBI theory but makes the
problem at hand much simpler to handle. One spin-off is that the description
of the physical behavior of various cosmological and stellar scenarios was
assumed to be controlled by the only remaining parameter $\kappa $. The
galactic halo is assumed to be filled with perfect fluid dark matter with
energy-momentum tensor $T^{\mu \nu }=pg^{\mu \nu }+(p+\rho )U^{\mu }U^{\nu }$%
, $g_{\mu \nu }U^{\mu }U^{\nu }=-1$ and the tangential velocity profile
provided by the Universal Rotation Curve (URC) [97,98]%
\begin{equation}
v_{\text{tg}}^{2}=v_{\infty }^{2}\frac{(r/r_{\text{opt}})^{2}}{(r/r_{\text{%
opt}})^{2}+r_{0}^{2}},
\end{equation}%
where $r_{\text{opt}}$ is the optical radius containing 83\% of the galactic
luminosity, $r_{0}$ is the halo core radius in units of $r_{\text{opt}}$,
the asymptotic velocity $v_{\infty }^{2}=v_{\text{opt}}^{2}(1-\beta _{\ast
})(1+r_{0}^{2})$, $v_{\text{opt}}=v_{\text{tg}}(r_{\text{opt}})$, $\beta
_{\ast }=0.72+0.44$Log$_{10}(L/L_{\ast })$, $L_{\ast }=10^{10.4}L_{\odot }$.
For spiral galaxies, $r_{0}=1.5(L/L_{\ast })^{1/5}$. Under the Newtonian
approximations that the pressure $p\simeq 0$, $8\pi \kappa \rho \ll 1$, and $%
(r/r_{\text{opt}})^{2}\gg r_{0}^{2}$, the EiBI field equations yield the
Lane-Emden equation with polytropic index $n=1$, which has an exact
nonsingular solution for dark matter density distribution as [95,96]  \footnote{The solution (3) 
to the Lane-Emden equation and its connection to dark matter was first pointed out in [96].}
\begin{equation}
\rho ^{(0)}(r)=\rho _{0}\left[ \frac{\sin \left( r\sqrt{\frac{2}{\kappa }}%
\right) }{r\sqrt{\frac{2}{\kappa }}}\right] ,
\end{equation}%
where $\rho ^{(0)}(0)=\rho _{0}$ is the constant central density. Assuming
that the halo has a sharp boundary $R_{\text{DM}}$, where the density
vanishes such that $\rho ^{(0)}(R_{\text{DM}})=0$, one has%
\begin{equation}
R_{\text{DM}}=\pi \sqrt{\frac{\kappa }{2}}.
\end{equation}%
The density profile (3) exhibits an unphysical behavior of becoming negative
for $R_{\text{DM}}<R<2R_{\text{DM}}$, which is why one has to require a
sharp halo boundary $\rho ^{(0)}(R\geq R_{\text{DM}})=0$.\footnote{%
We thank an anonymous referee for pointing this out.} This quite specific
behavior of the density profile differs from those of Navarro-Frenk-White
(NFW) or Burkert density profiles (that decay to zero only as $r\rightarrow
\infty $).\footnote{%
The observational situation is that the mass density profiles on the
logarithmic scale clearly indicate that the density is decaying at a finite
radius [104]. While NFW or Burkert density profiles might lead to different
physical metrics of interest, the key point of the present investigation is
that the dark matter has a finite radius following from \textit{stability}
considerations [100]. See also footnote 11.}

The mass profile of the dark matter is 
\begin{equation}
M(r)=4\pi \int_{0}^{r}\rho ^{(0)}(r)r^{2}dr=\frac{4R_{\text{DM}}^{3}}{\pi
^{2}}\rho _{0}\left[ \sin (\overline{r})-\overline{r}\cos (\overline{r})%
\right] ,
\end{equation}%
where the dimensionless quantity $\overline{r}=\pi r/R_{\text{DM}}$. The
total mass of the dark matter $M_{\text{DM}}$ and the mean density $%
\left\langle \rho \right\rangle $ in EiBI theory are given respectively by%
\begin{equation}
M_{\text{DM}}=M(R_{\text{DM}})=\frac{4}{\pi }\rho _{0}R_{\text{DM}}^{3}\text{%
, \ }\left\langle \rho \right\rangle =\frac{3\rho _{0}}{\pi ^{2}}.
\end{equation}%
The approximate physical metric has been derived in [35] as\footnote{%
The role of the URC profile (2) is that it determines the metric function $%
B(r)$ $\left[ \equiv e^{\nu (r)}\right] $. Following Chandrasekhar [105], we
know that the tangential velocity is given by $v_{\text{tg}}^{2}=\frac{r}{%
2e^{\nu }}(e^{\nu })_{,r}$. See the details in Ref.[93]. Using the URC for $%
v_{\text{tg}}^{2}$, defining $r/r_{\text{opt}}=x$, and integrating, we
obtain $e^{\nu (r)}=(x^{2}+r_{0}^{2})^{v_{\infty }^{2}}$. Restoring the
original variable $r/r_{\text{opt}}$, and defining $\overline{r}=\pi r/R_{%
\text{DM}}$, we immediately obtain the metric function $B(\overline{r})$ of
Eq.(8).} 
\begin{equation}
d\tau ^{2}=-B(\overline{r})dt^{2}+A(\overline{r})d\overline{r}^{2}+\overline{%
r}^{2}C(\overline{r})(d\theta ^{2}+\sin ^{2}\theta d\phi ^{2}),
\end{equation}%
\begin{equation}
B(\overline{r})=e^{\nu _{0}}\left[ \left( \frac{R_{\text{DM}}}{\pi r_{\text{%
opt}}}\right) ^{2}\overline{r}^{2}+r_{0}^{2}\right] ^{v_{\infty }^{2}},
\end{equation}%
\begin{equation}
A(\overline{r})=\left( \frac{R_{\text{DM}}}{\pi }\right) ^{2}\frac{1}{\left[
1-\frac{\overline{\rho }_{0}}{\overline{r}}\sin (\overline{r})+\overline{%
\rho }_{0}\cos (\overline{r})\right] \left[ 1-\frac{\overline{\rho }_{0}}{%
\overline{r}}\sin (\overline{r})\right] },
\end{equation}

\begin{equation}
C(\overline{r})=\left( \frac{R_{\text{DM}}}{\pi }\right) ^{2}\left[ 1-\frac{%
\overline{\rho }_{0}}{\overline{r}}\sin (\overline{r})\right] ,
\end{equation}%
where $e^{\nu _{0}}$ is an arbitrary constant of integration (which we set
to unity) and the dimensionless quantity $\overline{\rho }_{0}=\frac{8\rho
_{0}R_{\text{DM}}^{2}}{\pi }$.

Note that the surface area of a sphere at the boundary of dark matter halo
defined by $\overline{r}=\pi $, has the value $S=4\pi \overline{r}^{2}C(%
\overline{r})=4\pi \overline{r}^{2}\left( \frac{R_{\text{DM}}}{\pi }\right)
^{2}=4\pi R_{\text{DM}}^{2}$, which is just the spherical surface area in
"standard coordinates". Thus the dark matter radius $R_{\text{DM}}$ can be
identified with standard coordinate halo radius $R_{\text{HR}}$ derived
below. We shall need some of the above equations in the sequel.

\begin{center}
\textbf{II (b): MANNHEIM-O'BRIEN MODEL }
\end{center}

The unique Weyl action is 
\begin{eqnarray}
S_{\text{Weyl}} &=&-\alpha _{\text{g}}\int d^{4}x\sqrt{-g}\left[ C_{\lambda
\mu \nu \sigma }C^{\lambda \mu \nu \sigma }\right]  \notag \\
&=&-2\alpha _{\text{g}}\int d^{4}x\sqrt{-g}\left[ R_{\mu \sigma }R^{\mu
\sigma }-\frac{1}{3}\left( R_{\alpha }^{\alpha }\right) ^{2}\right] ,
\end{eqnarray}%
where $C^{\lambda \mu \nu \sigma }$ is the Weyl tensor and $\alpha _{\text{g}%
}$ is the dimensionless gravitational constant. The resulting field
equations are fourth order and trace free, rather long and complicated, so
we omit them here. The exact solution of vacuum\textit{\ }Weyl gravity for
the metric ansatz

\begin{equation}
d\tau ^{2}=-B(r)dt^{2}+\frac{1}{B(r)}dr^{2}+r^{2}(d\theta ^{2}+\sin
^{2}\theta d\phi ^{2})
\end{equation}%
was derived by Mannheim \& Kazanas [43] that describes the metric outside of
a localized static, spherically symmetric source of radius $r_{0}$ embedded
in a region with $T_{\mu \nu }(r>r_{0})=0$ as follows (after suitably
redefining the constants):%
\begin{equation}
B(r)=(1-6M\gamma )^{1/2}-\frac{2M}{r}+\gamma r-kr^{2},
\end{equation}%
where $M,\gamma ,k$ are constants of integration. Schwarzschild solution is
recovered at $\gamma =0,k=0$ as a special case of Weyl gravity.

On the other hand, in Refs.[48,107], the arguments and calculations instead
proceed from the considerations of \textit{potential}. In Weyl gravity, a
given local gravitational source generates a gravitational potential per
unit solar mass as follows (\textit{Eq.(8)} of Ref.[107]): 
\begin{equation*}
V^{\ast }=-\frac{\beta ^{\ast }c^{2}}{r}+\frac{\gamma ^{\ast }c^{2}r}{2},
\end{equation*}%
where $\beta ^{\ast }$ and $\gamma ^{\ast }$ are constants. Then, on
integrating $V^{\ast }$ over the local luminous matter distribution, one
obtains the local contribution to tangential velocity at $r=R>4R_{0}$ [107]:

\begin{equation}
\frac{v_{\text{loc}}^{2}}{R}\simeq \frac{N^{\ast }\beta ^{\ast }c^{2}}{R^{2}}%
\left( 1-\frac{9R_{0}^{2}}{2R^{2}}\right) +\frac{N^{\ast }\gamma ^{\ast
}c^{2}}{2}\left( 1-\frac{3R_{0}^{2}}{2R^{2}}-\frac{45R_{0}^{4}}{8R^{4}}%
\right) .
\end{equation}%
The meanings and values of the symbols above are as follows: $R_{0}$ is the
scale length such that most of the surface brightness is contained in $R\leq
4R_{0}$ of the optical disk region, $N^{\ast }$ is\ total number of solar
mass units in the luminous galaxy obtained via the mass-to-light ratio: $%
(M/L)L=M_{\text{lum}}=N^{\ast }M_{\odot }$. Thereafter, detailed arguments
(See Refs.[48,107]) are used to introduce two additional potentials of 
\textit{cosmologial} origin, viz., $\frac{\gamma _{0}c^{2}R}{2}$ and $%
-kc^{2}R^{2}$,\ that contribute to velocity such that%
\begin{equation}
\frac{v_{\text{tg}}^{2}}{R}=\frac{v_{\text{loc}}^{2}}{R}+\frac{\gamma
_{0}c^{2}}{2}-kc^{2}R.
\end{equation}%
The numerical values of the constants giving best fits to rotation curves of
all the 111 galaxy samples in [107] are:%
\begin{eqnarray}
\beta ^{\ast } &=&GM_{\odot }/c^{2}=1.48\times 10^{5}\text{ cm, }\gamma
^{\ast }=5.42\times 10^{-41}\text{ cm}^{-1}\text{,}  \notag \\
\gamma _{0} &=&3.06\times 10^{-30}\text{ cm}^{-1}\text{, }k=9.54\times
10^{-54}\text{ cm}^{-2}\text{.}
\end{eqnarray}%
\ It is evident from Eq.(15) that, in making the fits, the \textit{only }%
parameter that can vary from one galaxy to the other is the mass-to-light
ratio ($M/L$) leading to a galaxy dependent $N^{\ast }$. The mass of HI gas
is known and included in the fit. With everything else being universal, no
hypothetical dark matter is needed $-$ potential effects of cosmological
origin plus the local potential caused by the luminous mass of a galaxy are
enough to account for the observed rotation data.

For each galaxy with specific value of $N^{\ast }$ and other fixed constants
as in (16), $v_{\text{tg}}^{2}$ of Eq.(15) gives a \textit{finite} value of $%
R$ at the terminating velocity $v_{\text{tg}}^{2}(R_{\text{term}})=0$, where
the potentials balance. To illustrate various results derived here, we need
to consider samples and so in the following we choose the LSB galactic
sample ESO 1200211. The reason for this choice is that it is one of the
samples, whose central dark matter density has been recently fitted to
various known density profiles by Robles \& Matos [108] and thus it lends
itself to easy comparison with the results of the present paper. The plot in
Fig.1 shows that the rotation curve $v_{\text{tg}}^{2}$ decays to zero at a
radial distance $R_{\text{term}}=$ $52.04$ kpc, but this is still not the
halo radius! The actual halo radius $R_{\text{WR}}$, defined by the
stability inequality (21) below, will always be less than the value of $R_{%
\text{term}}$ for reasons of stability, as will soon be worked out.\footnote{%
Note that there are two specific profiles used in the paper: One is the URC
velocity profile in Eq.(2) that is never zero at any radius, $v_{\text{tg}%
}^{2}|_{\text{URC}}\neq 0$, while the other is the MO velocity profile,
Eq.(15), that can become zero at a finite radius $r=R_{\text{term}}$ such
that $v_{\text{tg}}^{2}|_{\text{MO}}(R_{\text{term}})=0.$ Such different
behavior\ might call into question the validity of the identification $R_{%
\text{DM}}=R_{\text{WR}}$ used as an input in this paper.
\par
The physical meaning of this input is explained on p.4. Here we point out
some additional grounds justifying the input. Note that it is quite logical
that two different theories can predict the \textit{same} measurable
quantity, say, light deflection. Similarly, despite differences in the
behavior, the profiles still provide fitted values of the same\textit{\ }%
quantity $\rho _{0}$. Ideally, the values should exactly be the same, which
is not the case, but they are comparable at least by order of magnitudes. In
the present case, it is the assumed equality of the radii $R_{\text{DM}}$
and $R_{\text{WR}}$ provided by the two profiles that is in question. We
have equated them\ on the ground that, observationally, there has to be only
one dark matter radius for each sample, no matter how it is defined. So the
equality in a way suggests itself. The other ground is that we have been
motivated by the approach originally proposed by Edery \& Paranjape [89],
where they equated the Einstein and Weyl angles. Actually, such
identifications are justified only \textit{a posteriori}, when they are
found to yield results that are independently known to be true. For
instance, using the input $\theta _{\text{E}}=\theta _{\text{W}}$ for the
deflection of light by galaxies, Edery \& Paranjape [89] obtained the value
of $\gamma $ in the metric (18) that is reasonably close to that obtained
independently in [48,107] by rotational data fits to samples. On the other
hand, we also know that Weyl and Einstein theories, giving respectively $%
\theta _{\text{W}}$ and $\theta _{\text{E}}$, are very different from each
other $-$ the former involves fourth order equations and is conformally
invariant, while the latter shares none of these. For this reason, at the
very outset in our paper, we clarified that we were not talking of merging
or mapping the two theories entirely into one another\textit{\ (p.4)} but
concentrating only on a particular common prediction.
\par
Similarly, despite differences in the EiBI theory and the MO model, our
identification $R_{\text{DM}}=R_{\text{WR}}$ leads to constraints on $\rho
_{0}$ that are found to be remarkably compatible with the independent NFW or
Burkert data fits $-$ that is, the fitted central density values do fall
within the stability induced limits. This is the \textit{a posteriori}
justification for using the identity $R_{\text{DM}}=R_{\text{WR}}$.}

The MO\ prescription for\ $v_{\text{tg}}^{2}$ in Eq.(15) leads to the
Mannheim-Kazanas metric (13) of Weyl gravity in the limit of large distances
away from the galactic core in which we are interested. It can be seen as
follows. The geodesic for a single test particle yields the tangential
velocity of material circular orbits at the arbitrary radius $r=R$ as [107]%
\begin{equation*}
v_{\text{tg}}^{2}=\left( Rc^{2}/2\right) B^{\prime },
\end{equation*}%
where prime denotes derivative with respect to $R$. Integrating, we obtain

\begin{eqnarray}
B(R) &=&1-\frac{2N^{\ast }\beta ^{\ast }}{R}+\left( N^{\ast }\gamma ^{\ast
}+\gamma _{0}\right) R-kR^{2}+  \notag \\
&&\frac{3R_{0}^{2}N^{\ast }\gamma ^{\ast }}{2R}+\frac{15R_{0}^{4}N^{\ast
}\gamma ^{\ast }-24R_{0}^{2}N^{\ast }\beta ^{\ast }}{8R^{3}}.
\end{eqnarray}%
Fit of $v_{\text{tg}}^{2}$ of Eq.(15) with the observed rotational data [107]
reveals that the constant $\gamma $ ($\equiv N^{\ast }\gamma ^{\ast }+\gamma
_{0}$) is of the order of $10^{-30}$ cm$^{-1}$ since $N^{\ast }\approx
10^{11}$ (roughly, the number of stellar units in a galactic luminous mass).
Also, the estimates covering all samples in [44] reveal that the luminous
masses lie approximately in the range $M=N^{\ast }\beta ^{\ast }\approx
10^{14}-10^{16}$ cm and from the fitted value of $\gamma ^{\ast }=5.42\times
10^{-41}$ cm$^{-1}$,\ it follows that $N^{\ast }\gamma ^{\ast }\approx
10^{-30}$ cm$^{-1}.$ Thus at halo distances ($R>4R_{0}$), we can ignore the
last two terms in (18) by order of magnitudes. In the same manner, we see
that $M\gamma <<1$ and it may be\ easily ignored in (13) so that $%
(1-6M\gamma )^{1/2}\simeq 1$. Thus the theoretical metric\ (13) and the
fitted metric (18) coincide at the form%
\begin{equation}
B(r)\simeq 1-\frac{2M}{r}+\gamma r-kr^{2}
\end{equation}%
at halo distances $R>4R_{0}$. This is also the form of\ the Weyl solution
used in Refs. [48,89]. Rephrasing, we can say that $v_{\text{tg}}^{2}$ can
be arrived at by differentiating the metric (17), which in turn approximates
to the Weyl solution (18). That's the relevance of the Weyl solution in the
fitting program.

In the asymptotic limit, Eq.(15) gives

\begin{equation}
\frac{v_{\text{tg}}^{2}}{R}\rightarrow \frac{N^{\ast }\beta ^{\ast }c^{2}}{%
R^{2}}+\frac{N^{\ast }\gamma ^{\ast }c^{2}}{2}+\frac{\gamma _{0}c^{2}}{2}%
-kc^{2}R,
\end{equation}%
in which one recognizes the Schwarzschild-like potential $V_{\beta ^{\ast
}}=N^{\ast }\beta ^{\ast }c^{2}/R$, two linear potential terms, viz., a
local $V_{\gamma ^{\ast }}=N^{\ast }\gamma ^{\ast }c^{2}R/2$ associated with
the matter distribution within a galaxy and a global $V_{\gamma _{0}}=\gamma
_{0}c^{2}R/2$ associated with the cosmological background, while the
universal de Sitter-like quadratic potential term $V_{k}=-kc^{2}R^{2}$ is
induced by inhomogeneities in the cosmic background. Note that the last
three potentials are new inputs into the MO model [48] designed to interpret
the rotation curve data.

The radial geodesic in the metric (12) is given by%
\begin{equation}
\left( \frac{dr}{dt}\right) ^{2}=B^{2}(r)-a\frac{B^{3}(r)}{r^{2}}-bB^{3}(r),
\end{equation}%
where $a$ \ and $b$ are constants of motion. The right hand side of the
above equation is the "effective potential" $V$, and $V_{\text{MO}}^{\prime
\prime }\equiv f(R)$ involves the derivatives of $B(R)$ that, in turn,
contains the so called quadratic potential $V_{k}$ ($=-kc^{2}R^{2}$)
introduced in Ref.[48]. [The subscript "MO" is used here to distinguish it
from the potential $V_{\text{EiBI}}$ to be defined in Eq.(22)]. This
potential $V_{k}$ is responsible for providing a finite radius $R_{\text{WR}%
} $.

The main reason is the \textit{negative sign} in $V_{k}$ needed for the data
fit by MO. Because of this, it is quite evident from\ the plot of $V_{\text{%
MO}}^{\prime \prime }$ [Fig.(2)] that the sample ESO 1200211 has a maximally
allowed finite halo radius\ $\sim 39.033$ kpc (see also [100] for similar
plots). On the other hand, if $V_{k}$ \ has a positive sign (in which case
no fit to data) or is altogether removed from $B(R)$, hence from $V_{\text{MO%
}}^{\prime \prime }$, it can be graphically verified that there will be%
\textit{\ no} finite stable radius $R_{\text{WR}}$ for the halo. This shows
the crucial role of $V_{k}$. Thus, it is the requirement of fitting to data
that indirectly limits the halo size to $R_{\text{WR}}$. As to the physical
reason, we see that\ the repulsive potential $V_{k}$ balances the remaining
attractive potentials at $r=R_{\text{term}}$ [see Eq.(15)] but stability
further demands that $R_{\text{WR}}<R_{\text{term}}$, as is evident from
Figs.1 \& 2. Therefore, we can say that, at $r=R_{\text{WR}}$, attractive
potentials prevail over the repulsive potential $V_{k}$ constraining the gas
on the circular orbit, as it should.

The right hand side of Eq.(20) and its first derivative with respect to $r$,
both must vanish at the circular radius $r=R$ giving 
\begin{equation*}
a=\frac{R^{3}B^{\prime }(R)}{2B^{2}(R)}\text{, \ \ }b=\frac{2B(R)-RB^{\prime
}(R)}{2B^{2}(R)}.
\end{equation*}%
The condition for stability is that the second derivative of the
\textquotedblleft effective potential\textquotedblright\ $V$ \ with respect
to $r$ must be negative at the circular radius $r=R$. The resultant
expression with values of $a,b$ plugged in leads to the generic stability
criterion for the MO model: 
\begin{equation}
V_{\text{MO}}^{\prime \prime }=2B^{\prime 2}(R)-B(R)B^{\prime \prime
}(R)-3B(R)B^{\prime }(R)/R<0.
\end{equation}%
This inequality graphically predicts a \textit{finite, stable, maximum} halo
radius that we call $R_{\text{WR}}$ caused solely by the quadratic potential 
$V_{k}(R)=-kc^{2}R^{2}$. Interestingly, the predicted $R_{\text{WR}}$ lends
itself to observational testing in the near future as its value does not
often much exceed the $R_{\text{last}}$ for many samples.

We shall apply the above algorithm to many samples but for illustration, we
display $V_{\text{MO}}^{\prime \prime }$ \textit{vs} $R$ for the same sample
ESO 1200211 in Fig.2. The value of $N^{\ast }$ can be found from $N^{\ast
}=M_{\text{lum}}/\beta ^{\ast }$, where $M_{\text{lum}}=[(M/L)_{\text{stars}%
}\times L_{\text{B}}$ $+M_{\text{HI}}]\times 10^{10}M_{\odot }$. All
necessary components can be read off from the entries in the \textit{Table IV%
} in [107]. The value of $N^{\ast }$ ($=5.60\times 10^{7}$) together with
other constants in (15), when plugged into the inequality (21), immediately
graphically yields $R=R_{\text{WR}}=39.03$ kpc, which is less than $R_{\text{%
term}}$ calculated above.

\begin{center}
\textbf{III. UPPER LIMIT DENSITY FOR STABILITY}
\end{center}

To analyze the stability of circular orbits, one needs to analyze the second
order derivative of the concerned potential, which we wish to do here. To
find the potential $V$, note that the four velocity $U^{\alpha }=\frac{%
dx^{\sigma }}{d\tau }$ of a test particle of rest mass $m_{0}$ moving in the
halo (restricting ourselves to $\theta =\pi /2$) follows the equation $%
g_{\nu \sigma }U^{\nu }U^{\sigma }=-m_{0}^{2}$ that can be cast into a
Newtonian form in the dimensionless radial variable $\overline{r}$ ($=\pi
r/R_{\text{DM}}$) as 
\begin{equation}
\left( \frac{d\overline{r}}{d\tau }\right) ^{2}=E^{2}+V_{\text{EiBI}}(%
\overline{r})
\end{equation}%
which gives, for the metric Eqs.(7)-(10) of Sec.II(a), the EiBI potential 
\begin{equation}
V_{\text{EiBI}}(\overline{r})=\left[ E^{2}\left\{ \frac{1}{AB}-1\right\} -%
\frac{L^{2}}{AC\overline{r}^{2}}-\frac{1}{A}\right]
\end{equation}%
\begin{equation}
E=\frac{U_{0}}{m_{0}},L=\frac{U_{3}}{m_{0}},
\end{equation}%
where the constants $E$ and $L$, respectively, are the conserved
relativistic energy and angular momentum per unit mass of the test particle.
Circular orbits at any arbitrary radius are defined by $\overline{r}=%
\overline{R}=$ constant, so that $\frac{d\overline{r}}{d\tau }\mid _{%
\overline{r}=\overline{R}}=0$ and, additionally, $\frac{dV}{d\overline{r}}%
\mid _{\overline{r}=\overline{R}}=0$. From these two conditions follow the
conserved quantities as under: 
\begin{equation}
L^{2}=\frac{X}{Z}
\end{equation}%
and using it in $V_{\text{EiBI}}(\overline{R})=-E^{2}$, we get 
\begin{equation}
E^{2}=\frac{Y}{Z},
\end{equation}%
where%
\begin{equation}
X\equiv -\kappa ^{2}\overline{R}^{3}v_{\infty }^{2}(\overline{R}-\overline{%
\rho }_{0}\sin \overline{R})^{2}
\end{equation}%
\begin{equation}
Y\equiv \left( \kappa \overline{R}^{2}+2r_{0}^{2}r_{\text{opt}}^{2}\right)
\left( r_{0}^{2}+\frac{\kappa \overline{R}^{2}}{2r_{\text{opt}}^{2}}\right)
^{v_{\infty }^{2}}\left( \overline{\rho }_{0}\overline{R}\cos \overline{R}+%
\overline{\rho }_{0}\sin \overline{R}-2\overline{R}\right)
\end{equation}%
\begin{eqnarray}
Z &\equiv &\left\{ \kappa \overline{R}^{2}\left( 1-2v_{\infty }^{2}\right)
+2r_{0}^{2}r_{\text{opt}}^{2}\right\} \overline{\rho }_{0}\sin \overline{R}%
+\left( \kappa \overline{R}^{2}+2r_{0}^{2}r_{\text{opt}}^{2}\right) 
\overline{\rho }_{0}\overline{R}\cos \overline{R}  \notag \\
&&-4\overline{R}r_{0}^{2}r_{\text{opt}}^{2}-2\kappa \overline{R}^{3}\left(
1-v_{\infty }^{2}\right) .
\end{eqnarray}

Putting the expressions for $L^{2}$ and $E^{2}$ in Eq.(23), we find the
complete expression for $V_{\text{EiBI}}$. The orbits will be stable if $V_{%
\text{EiBI}}^{\prime \prime }\equiv $ $\frac{d^{2}V}{d\overline{r}^{2}}\mid
_{\overline{r}=\overline{R}}<0$ and unstable if $V_{\text{EiBI}}^{\prime
\prime }>0$. The expression for $V_{\text{EiBI}}^{\prime \prime }$ is

\begin{eqnarray}
&&V_{\text{EiBI}}^{\prime \prime }\left( \overline{R};\kappa ,\overline{\rho 
}_{0},r_{0},r_{\text{opt}},v_{\infty }\right)  \notag \\
&=&\left[ \frac{2v_{\infty }^{2}\left( \overline{R}+\overline{\rho }_{0}%
\overline{R}\cos \overline{R}-\overline{\rho }_{0}\sin \overline{R}\right) }{%
\overline{R}^{2}\left( \kappa \overline{R}^{2}+2r_{0}^{2}r_{\text{opt}%
}^{2}\right) Z}\right] \times  \notag \\
&&\left[ 32\overline{R}^{2}r_{0}^{2}r_{\text{opt}}^{2}+8\kappa \overline{R}%
^{4}\left( 1-v_{\infty }^{2}\right) +6\overline{\rho }_{0}^{2}r_{0}^{2}r_{%
\text{opt}}^{2}\left( 1+\overline{R}^{2}\right) \right.  \notag \\
&&\left. +\kappa \overline{\rho }_{0}^{2}\overline{R}^{2}\left( 1-2v_{\infty
}^{2}+3\overline{R}^{2}\right) \right.  \notag \\
&&\left.-2\overline{R}^{2}\left\{ 14r_{0}^{2}r_{\text{opt}}^{2}+\kappa 
\overline{R}^{2}\left( 5-2v_{\infty }^{2}\right) \right\} \overline{\rho }%
_{0}\cos \overline{R}\right.  \notag \\
&&\left. +\left\{ 2\left( \overline{R}^{2}-3\right) r_{0}^{2}r_{\text{opt}%
}^{2}-\kappa \overline{R}^{2}\left( 1-2v_{\infty }^{2}-\overline{R}%
^{2}\right) \right\} \overline{\rho }_{0}^{2}\cos \left( 2\overline{R}%
\right) \right.  \notag \\
&&\left. -\left\{ 36r_{0}^{2}r_{\text{opt}}^{2}+4\overline{R}^{2}r_{0}^{2}r_{%
\text{opt}}^{2}+6\kappa \overline{R}^{2}+2\kappa \overline{R}^{4}-12\kappa 
\overline{R}^{2}v_{\infty }^{2}\right\} \overline{\rho }_{0}\overline{R}\sin 
\overline{R}\right.  \notag \\
&&\left. +\left\{ 6r_{0}^{2}r_{\text{opt}}^{2}+\kappa \overline{R}%
^{2}-2\kappa \overline{R}^{2}v_{\infty }^{2}\right\} \overline{\rho }_{0}^{2}%
\overline{R}\sin \left( 2\overline{R}\right) \right] .
\end{eqnarray}

From the above expression, it is absurd to straightforwardly draw any
conclusion about stability or otherwise of the circular orbits. Clearly,
much will depend on the parameter ranges chosen on the basis of physical
considerations. While other parameters can be reasonably assigned, the as
yet unknown parameters are the dark matter radius $\kappa $ ($=2R_{\text{DM}%
}^{2}/\pi ^{2}$) and the dimensionless central density $\overline{\rho }_{0}$
($=8\rho _{0}R_{\text{DM}}^{2}/\pi $), again depending only on $\kappa $. In
the first order approximation, the density distribution in the dark matter
has been assumed in [95,96] to be low such that $8\pi G\kappa \rho
^{(0)}/c^{4}<<1$, but the central density $\rho _{0}$ could still be large
since $\left\vert \sin (x)/x\right\vert \leq 1$ [see Eq.(3)]. The question
therefore is how large or small could it be, or turning it around, could
there be any upper limit on $\rho _{0}$ imposed by the stability criterion?

The answer is in the affirmative and can be found graphically. We find that $%
V^{\prime \prime }$ is indeed very sensitive to changes in $\rho _{0}$
leading to different upper limits $\rho _{0}^{\text{upper}}$ for different
galactic samples such that stable circular orbits are possible only when $%
\rho _{0}\leq \rho _{0}^{\text{upper}}$. The reason is that $R_{\text{DM}}$
changes from sample to sample, as it should, and thereby leads to different
\ (though not too different) values for $\kappa $ and $\rho _{0}^{\text{upper%
}}$. Let us again consider the previous sample ESO 1200211, a low surface
brightness galaxy with a halo/dark matter radius $R_{\text{WR}}\equiv R_{%
\text{DM}}=39.03$ kpc that corresponds to $\kappa =308.74$ kpc$^{2}$. With $%
\kappa $ thus fixed, we fix other parameters respecting the Newtonian
approximation\footnote{%
The range of $r$ and $r_{\text{opt}}$ is chosen so as to ensure the
Newtonian approximation $r/r_{\text{opt}}\gg r_{0}$, while the approximate
value of $v_{\infty }^{2}$ is an observed fact. The formula for $r_{0}$ for
spiral galaxies is $r_{0}=1.5\times (L_{\text{B}}/10^{10.4}L_{\odot }$)$%
^{1/5}$, which evaluates to $r_{0}=0.61$ for the sample ESO 1200211, where $%
r_{\text{opt}}\simeq 4R_{0}$, $L_{\text{B}}=0.028\times 10^{10}L_{\odot }$, $%
R_{0}=2$ kpc. Data taken from [107].
\par
\bigskip}, e.g., $r_{0}=0.61$, $v_{\infty }^{2}=0.000001$, $r_{\text{opt}}=8 
$ kpc, with the dimensionless radius $\overline{R}$ $(=\pi R/R_{\text{DM}})$
chosen in the range $\overline{R}$ $\in \lbrack 0.5\pi ,\pi ]$ corresponding
to coordinate radii $R\in \lbrack 19.52$ kpc, $39.03$ kpc$]$, the
dimensionless density parameter chosen in the range $\overline{\rho }_{0}\in
\lbrack 0.25\pi ,$ $0.8\pi ]$ and plot $V_{\text{EiBI}}^{\prime \prime }$ 
\textit{vs} $\overline{R}$ using the expression (30).

Graphical analysis shows that, while $V_{\text{EiBI}}^{\prime \prime }$ is
not much sensitive to the variation of the other parameters within the
Newtonian approximation, it is \textit{greatly} sensitive to the variation
of the remaining parameter $\rho _{0}$. Figs. 3 and 4 respectively show
that, for values of $\overline{\rho }_{0}>0.94$, there is instability in the
entire or partial range of the halo radii $\overline{R}$, while Fig. 5 tells
us that there is an upper limit occurring at $\overline{\rho }_{0}^{\text{%
upper}}=$ $0.94=\lambda ^{\text{upper}}\pi $, where $\lambda ^{\text{upper}%
}=0.299$, such that for $\overline{\rho }_{0}\leq \overline{\rho }_{0}^{%
\text{upper}}$, all circular orbits in the entire chosen radial range for $%
\overline{R}$ are stable. It can be verified that this value of $\lambda ^{%
\text{upper}}$ surprisingly remains the \textit{same} for values for $\kappa 
$ across the entire range of 111 samples (some tabulated here), so $\rho
_{0}^{\text{upper}}$ is quite a reliable limit.

Rewriting in terms of $\rho _{0}$, we have%
\begin{equation}
\rho _{0}^{\text{upper}}=\frac{\overline{\rho }_{0}^{\text{upper}}\pi }{8R_{%
\text{DM}}^{2}}=\frac{\lambda ^{\text{upper}}}{4\kappa }.
\end{equation}%
This by itself is an interesting prediction of EiBI theory. However, if we
want to quantify $\rho _{0}^{\text{upper}}$ for a given galaxy, we need to
use the value of $R_{\text{DM}}$ but since concrete observed values are yet
unavailable, we choose to use the input $R_{\text{DM}}=R_{\text{WR}}$. It
turns out that this choice, though not mandatory, works well giving
definitive values for $\rho _{0}^{\text{upper}}$ for all samples. Plugging
in the values of $\lambda ^{\text{upper}}$ and $\kappa $, we find that the
constraint $\overline{\rho }_{0}\leq \overline{\rho }_{0}^{\text{upper}}$
immediately translates into a generic constraint such that for 
\begin{equation}
\rho _{0}\leq \rho _{0}^{\text{upper}}\text{,}
\end{equation}%
all circular orbits in the chosen range for $R$ are stable. Thus, using the
value of $\kappa $ as above in Eq.(31), the sample ESO 1200211
quantitatively yields $\rho _{0}^{\text{upper}}=5.04\times 10^{12}$ $%
M_{\odot }$kpc$^{-3}$. In general, as long as $\rho _{0}$ of any galaxy
obeys the stability induced constraint (32), the circular material orbits in
the halo region will be stable up to a \textit{maximum} radius $R=R_{\text{WR%
}}$.

\begin{center}
\textbf{IV. CENTRAL AND MEAN DARK MATTER DENSITY}
\end{center}

So far, the algorithm has been as follows: Take any galactic sample, find $%
R_{\text{WR}}$ for that sample using the method of Sec.II(b). Then, from the
identity $R_{\text{DM}}=R_{\text{WR}}$, find $R_{\text{DM}}$ (hence $\kappa $%
) and using Eq.(31), find $\rho _{0}^{\text{upper}}$. However, we still do
not know the values of $\rho _{0}$ for all the samples observed to date and
cannot ascertain whether or not they satisfy the stability induced
constraint $\rho _{0}\leq \rho _{0}^{\text{upper}}$. On the other hand, some
notable dark matter simulations and profiles for several samples show values
for $\rho _{0}$ that \textit{do} respect this constraint (Table II). This
success then prompts us to ask if there is any lower limit on $\rho _{0}$
such that $\rho _{0}^{\text{lower}}\leq \rho _{0}\leq \rho _{0}^{\text{upper}%
}$ holds.

Fortunately, there is a way to find the values of $\rho _{0}^{\text{lower}}$%
, once we are able to estimate the total mass of dark matter $M_{\text{DM}}$
using the observed mass-to-light ratios. Fitted data are available for the
luminous mass-to-light ratios ($M_{\text{lum}}/L_{\text{B}}$) in solar units 
$\Upsilon _{\odot }$ $(\equiv \frac{M_{\odot }}{L_{\odot }})$:

\begin{equation*}
\frac{M_{\text{lum}}}{L_{\text{B}}}=\gamma \Upsilon _{\odot }.
\end{equation*}%
The luminous mass ($M_{\text{lum}}$) of a galaxy is contributed mostly by
stars and gases excluding dark matter. The stellar mass-to-light ratios $%
\gamma $ for 111 samples \ in [107] are seen to lie between $0.2$ and $8$
(with just a couple of exceptions), which is consistent with the upper bound
of $10\Upsilon _{\odot }$ suggested by the population synthesis models [109].
On the other hand, there is no detectable dark matter associated with the
galactic disk, most of the dark matter is distributed in the halo [110-114].
So we can write

\begin{equation}
M_{\text{tot}}=M_{\text{lum}}+M_{\text{DM}},
\end{equation}%
then 
\begin{equation*}
\frac{M_{\text{tot}}}{L_{\text{B}}}=\beta \Upsilon _{\odot }
\end{equation*}%
where $\beta $ must be larger than $\gamma $, if there is dark matter
(observed mass-to-light ratios are still uncertain). We can\ thus write,
following Edery \& Paranjape [89]:

\begin{equation}
\frac{M_{\text{lum}}}{M_{\text{tot}}}=\frac{1}{\alpha },
\end{equation}%
which gives $\beta =\alpha \gamma $ and $\alpha $ should be so chosen as to
make $\beta >\gamma $. In general, one takes $\alpha >1$ such that $M_{\text{%
tot}}>M_{\text{lum}}$ thereby accommodating the presence of dark matter.

Assuming that the halos must be substantially larger than the last measured
point $R_{\text{last}}$, the dark to luminous mass within $R_{\text{last}}$
then gives an upper limit through $f_{b}<\frac{M_{\text{lum}}}{M_{\text{tot}}%
}$ and therefore $f_{b}<\left( 1+\frac{M_{\text{DM}}}{M_{\text{lum}}}\right)
^{-1}$. For some galaxies, $f_{b}<0.08$, as reported in de Blok and McGaugh
[115]. Thus, using (34), we have $f_{b}<\frac{1}{\alpha }$ and $f_{b}<0.08$.
Certainly, these inequalities do not constrain $\alpha $ in any way. One of
the infinity of options to ensure that both hold simultaneously is to assume
that $\frac{1}{\alpha }=0.08\Rightarrow \alpha =12.5$. While $\alpha $ can
be varied at will unless it is definitively fixed by independent concrete
observed data, we shall for the moment choose the value $\alpha =12.5$ only
to have an idea of the order of magnitudes of the estimated densities, but
we shall change $\alpha $ later. The current choice would imply $\beta \in
\lbrack 2.5,100]$ corresponding to $\gamma \in \lbrack 0.2,8]$. The values
of $\beta $ $\sim 100$ is enough to account for the large dark matter
content of LSB galaxies (i.e., large mass-to-light ratios $\frac{M_{\text{tot%
}}}{L_{\text{B}}}$) such as DDO154. Currently favored Burkert density
profile can provide an excellent mass model for the dark halos around disk
systems up to $100$ times more massive than small dwarf galaxies for which
the profile was initially intended [116,117].

The ratio $M_{\text{lum}}/M_{\text{tot}}$ then gives the connection between $%
M_{\text{DM}}$ of EiBI theory and the luminous mass $M_{\text{lum}}$ of
galaxies via Eq.(33): 
\begin{equation}
M_{\text{DM}}=(\alpha -1)M_{\text{lum}}\text{.}
\end{equation}%
Using $R_{\text{WR}}\equiv R_{\text{DM}}$, Eq.(6) can be rewritten as 
\begin{equation}
\rho _{0}^{\text{lower}}=\frac{\pi (\alpha -1)M_{\text{lum}}}{4R_{\text{WR}%
}^{3}},\text{ \ \ }\left\langle \rho \right\rangle ^{\text{lower}}=\frac{%
3\rho _{0}^{\text{lower}}}{\pi ^{2}}.
\end{equation}%
The superscript "lower" indicates that it is the lower limit of the dark
matter central density $\rho _{0}$ because $R_{\text{WR}}$ is the maximum
allowed halo radius (see Fig.2), where $V^{\prime \prime }<0$ gives stable
radii $R\leq R_{\text{WR}}$. Evidently, $\rho _{0}^{\text{lower}}$ is
proportional to the as yet unknown parameter $\alpha $. We are free to raise
the value of $\alpha $ arbitrarily, but then the consequent larger values of 
$\beta $ would lead to too large an amount of dark matter comparable to that
existing in galactic clusters.\footnote{%
At much larger scales of galactic clusters, the value of $\beta $ could be $%
\geq 120$ [118,119] so that $\frac{M_{\text{tot}}}{L_{\text{B}}}\geq 120$ in
solar units. We are not contemplating galactic clusters here.}

To illustrate the order of magnitudes involved for $\rho _{0}^{\text{lower}}$
, hence for\ $\left\langle \rho \right\rangle ^{\text{lower}}$, we again
consider the low mass LSB sample ESO 1200211 for which $M_{\text{lum}%
}=5.60\times 10^{7}$ $M_{\odot }$, $\gamma =0.2$ and our method in Sec.II(b)
yields $R_{\text{WR}}=39.03$ kpc, $\kappa =308.74$ kpc$^{2}$, $R_{\text{term}%
}=52.04$ kpc. Using Eqs.(36), and allowing for a fairly large amount of dark
matter over luminous matter corresponding to $\alpha =12.5$, we find $\rho
_{0}^{\text{lower}}=8.49\times 10^{4}$ $M_{\odot }$kpc$^{-3}$ and \ $%
\left\langle \rho \right\rangle ^{\text{lower}}=2.58\times 10^{4}$ $M_{\odot
}$kpc$^{-3}$. To compare these values of density, we consider several dark
matter density profiles: (i) the Bose-Einstein [108] condensate ($\rho _{0}^{%
\text{\textit{BEC}}}$), (ii) Pseudo-Isothermal [120] profile ($\rho _{0}^{%
\text{\textit{PI}}}$), (iii) the NFW [121,122] profile ($\rho _{i}^{\text{\textit{%
NFW}}}$), (iv) Burkert [116] profile ($\rho ^{\text{\textit{BP}}}$), (see
Sec.V below for the exact forms). They yield values as follows: (i) $\rho
_{0}^{\text{\textit{BEC}}}=1.38\times 10^{7}$ $M_{\odot }$kpc$^{-3}$, (ii) $%
\rho _{0}^{\text{\textit{PI}}}=4.64\times 10^{7}$ $M_{\odot }$kpc$^{-3}$ and
(iii) $\rho _{0}^{\text{\textit{NFW}}}=2.45\times 10^{6}$ $M_{\odot }$kpc$%
^{-3}$ for ESO 1200211 (see the entries in \textit{Table I }of [108]). In
Sec.III, we already found for the present sample the value $\rho _{0}^{\text{%
upper}}=5.04\times 10^{12}$ $M_{\odot }$kpc$^{-3}$. We hence see that\ the\
central densities from different profiles fall within the stability induced
limits, that is, $\rho _{0}^{\text{lower}}<\rho _{0}^{\text{\textit{BEC, PI
or NFW}}}<\rho _{0}^{\text{upper}}$ holds.\footnote{%
Once again, a question of compatibility might be phrased as follows: The
EiBI density\ profile has $\rho (R_{\text{DM}})=0$ and remains zero beyond $%
r>R_{\text{DM}}$, while, in contrast, the NFW and Burkert profiles have $%
\rho (r\rightarrow \infty )=0$. Since the asymptotic behavior of latter
density distributions are different from that of EiBI, determining whether
the data obtained by fitting to NFW or Burkert profiles fall within the
stability induced limits from EiBI theory calls into question the issue of
compatibility of the EiBI with those profiles.
\par
We wish to clarify\ that it is the central density $\rho _{0}$, a parameter
distinctly appearing in all density distributions, that is under present
investigation. No matter what the profile is, the target is always the same:
to find information about $\rho _{0}$ for any given galactic sample. We know
that NFW profile is cuspy ($\rho ^{\text{NFW}}\propto r^{-1}$), while others
such as that of Burkert are cored ($\rho ^{\text{BP}}\propto r^{0}$).
Despite this radical difference in behavior\textit{\ at the origin}, both
profiles are quite well accepted though \textit{per se} they are different.
One could rephrase this difference as incompatibility. The main thing
however is that the fitted values of central density from the two profiles
should approximately be the same, at least by order of magnitude, which
actually is the case [108,132]. (A brief account of comparisons\ as to which
profile fits the data better is given at the very last paragraph of our
paper). In the same vein, despite differences in the \textit{asymptotic }%
behavior of EiBI profile and other density profiles, the information about
the common parameter $\rho _{0}$ should approximately be the same. It would
be fair to say that the derived bound for the central density $\rho _{0}$ is
indeed in agreement with the astrophysical data but that it is yet to be
determined if the EiBI density profile is compatible with future N-body
simulation featuring higher accuracy allowing for testing the outskirts of
dark matter halos.
\par
Also, it is very unlikely that the attractive dark matter is spread all the
way to infinity, where repulsive dark energy takes over. The finite extent
of dark matter is supported by the observed rapid decline of velocity
dispersion after a certain radius [143]. It could indeed be interesting to
take this fact as an empirical input for $R_{\text{DM}}$ only if we knew the
exact radius at which the decline ended. Pending this knowledge, we used
theoretical inputs $R_{\text{WR}}$ for $R_{\text{DM}}$, which\ yielded the
interval for $\rho _{0}$ confirmed by data fits so far.} If we take $\alpha
<12.5$, which should also be quite acceptable for many samples, the values
of $\rho _{0}^{\text{lower}}$ will only be further lowered and of course the
interval will be well supported.

If we increase $\alpha $ to an (unlikely) mammoth value, say $\alpha =300$
so that $M_{\text{tot}}=300M_{\text{lum}}$ implying $\beta =60$ so that $%
\frac{M_{\text{tot}}}{L_{\text{B}}}=60\Upsilon _{\odot }$ in the considered
sample, then $\rho _{0}^{\text{lower}}\sim 2.20\times 10^{6}$ $M_{\odot }$kpc%
$^{-3}$, and we notice that the proposed limits are still not violated! If
we exclude NFW profile, then the values $\alpha $ can be increased even
further. This testifies to the validity of Eq.(36) as well as the limits.

\begin{center}
\textbf{V. Milky Way}
\end{center}

As for our Milky Way galaxy, the latest reported estimates are the
following: Using the gas terminal velocity curve, Sgr A$^{\ast }$ proper
motion, an oblate bulge + Miyamoto-Nagai disc and NFW halo, Kafle \textit{et
al.} [123] estimated the luminous (disc + bulge) mass to be $M_{\text{lum}%
}\sim 1.04\times 10^{11}M_{\odot }$, so that $N^{\ast }=1.04\times 10^{11}$
and the virial mass inclusive of dark matter $M_{\text{vir}}=M_{\text{tot}}$ 
$\sim 0.80\times 10^{12}$ $M_{\odot }$ so that $\alpha =M_{\text{tot}}/M_{%
\text{lum}}=7.68$. (We have not considered the data/fit uncertainties). For
alternative but not too different values of $M_{\text{tot}}$, see [124-131].%
\footnote{%
In [124], it is reported that $M_{\text{tot}}$ $=(1.4\pm 03)\times 10^{12}$ $%
M_{\odot }$ from tidal effects on globular clusters and $M_{\text{tot}}$ $%
=(1.4\pm 08)\times 10^{12}$ $M_{\odot }$ from globular cluster radial
velocities. The remarkable similarity between two completely independent
determinations of mass may be taken as a strong empirical signature for the
existence of dark matter around the Milky Way. In [125], the reported
estimate is $M_{\text{tot}}$ $=(1.0-1.5)\times 10^{12}$ $M_{\odot }$, again
not too different.}
Using the scale length $R_{0}=4.9$ kpc [123] and the
above $N^{\ast }$ in the inequality (21), we find that $R_{\text{WR}}=R_{%
\text{DM}}=111.90$ kpc. With this value of $R_{\text{WR}}$, Eq. (31) then
yields a value $\rho _{0\text{, MW}}^{\text{upper}}=6.14\times 10^{11}$ $%
M_{\odot }$kpc$^{-3}$ characteristic of the Milky way, which does not exceed
the maximum density ($\sim 10^{12}M_{\odot }$kpc$^{-3}$) proposed in this
paper.

The density profiles considered here are of the forms:%
\begin{eqnarray}
\rho ^{\text{\textit{BP}}}(r) &=&\frac{\rho _{0}^{\text{\textit{BP}}%
}r_{0}^{3}}{(r+r_{0})(r^{2}+r_{0}^{2})}=\rho _{0}^{\text{\textit{BP}}}\left(
1-\frac{r}{r_{0}}+\frac{r^{4}}{r_{0}^{4}}+...\right) \text{ \ \ \ [116]} \\
\rho ^{\text{\textit{NFW}}}(r) &=&\frac{\rho _{0}^{\text{\textit{NFW}}}}{%
(r/r_{S})[1+(r/r_{S})^{2}]}=
\rho _{0}^{\text{\textit{NFW}}}\left( \frac{r_{S}%
}{r}-\frac{r}{r_{S}}+\frac{r^{3}}{r_{S}^{3}}+...\right)  \notag \\ \text{[121,122]} \\
\rho ^{\text{\textit{PI}}}(r) &=&\frac{\rho _{0}^{\text{\textit{PI}}}}{%
1+(r/r_{c})^{2}}=\rho _{0}^{\text{\textit{PI}}}\left( 1-\frac{r^{2}}{%
r_{c}^{2}}+\frac{r^{4}}{r_{c}^{4}}+...\right) \text{ \ \ [120]} \\
\rho ^{\text{\textit{BEC}}}(r) &=&\rho _{0}^{\text{\textit{BEC}}}\left[ 
\frac{\sin \left( \xi r\right) }{\xi r}\right] =\rho _{0}^{\text{\textit{BEC}%
}}\left( 1-\frac{\xi ^{2}r^{2}}{6}+\frac{\xi ^{4}r^{4}}{120}+...\right) \notag \\
\text{[108]} 
\end{eqnarray}%
where $\rho _{0}$ is the galactocentric density, $\rho _{0}^{\text{\textit{%
NFW}}}$ is related to the density of the universe at the moment the halo
collapsed and $r_{0}$, $r_{S}$, $r_{c}$ are core, scale, characteristic
radii respectively, while $\xi =\sqrt{Gm^{3}/\hslash ^{2}a}$ in which $m$ is
the mass of the dark matter particle and $a$ is the scattering length [108].
The fitted latest data on Milky Way dark matter central density are $\rho
_{0}^{\text{\textit{BP}}}=4.13\times 10^{7}$ $M_{\odot }$kpc$^{-3}$ (Burkert
profile) and $\rho _{0}^{\text{\textit{NFW}}}=1.40\times 10^{7}$ $M_{\odot }$%
kpc$^{-3}$ (NFW profile) [132]. In both cases, we see that these fitted
values are four orders of magnitude\textit{\ less} than $\rho _{0\text{,MW}%
}^{\text{upper}}$, as desired. \ From Eq.(36), for $\alpha \sim 7.68$, we
get $\rho _{0\text{,MW}}^{\text{lower}}\sim 2.09\times 10^{3}$ $M_{\odot }$%
kpc$^{-3}$ and $\left\langle \rho \right\rangle _{0,\text{MW}}^{\text{lower}%
}\sim 6.37\times 10^{2}$ $M_{\odot }$kpc$^{-3}$ . Once again, we see that
the densities satisfy $\rho _{0}^{\text{lower}}<\rho _{0}^{\text{\textit{BP,
NFW}}}<\rho _{0}^{\text{upper}}$. \ A larger value of $\alpha $ does not
disallow the inequalities at either end.

Apart from the galactocentric density $\rho _{0}$, the local (at $R_{\odot
}=8.5$ kpc) dark matter density $\rho _{\odot }$ provides a strong basis for
the experimental endeavors for indirect detection of the dark matter. Though
there is broad consensus, different groups have come up with somewhat
different conclusions regarding the local density of dark matter. For
example, Kuijken and Gilmore [111-114] find a volume density near Earth $\rho
_{\oplus }\simeq 0.01$ GeV/cm$^{3}=2.64\times 10^{5}$ $M_{\odot }$kpc$^{-3}$%
. Other reported values are the following. Bahcall \textit{et al. }[133] find
a best-fit value of $\rho _{\odot }=0.34$ GeV/cm$^{3}=8.96\times 10^{6}$ $%
M_{\odot }$kpc$^{-3}$, Caldwell and Ostriker [134] find $\rho _{\odot }=0.23$
GeV/cm$^{3}=6.07\times 10^{6}$ $M_{\odot }$kpc$^{-3}$, while Turner [135]
calculates $\rho _{\odot }=0.3-0.6$ GeV/cm$^{3}=7.92\times 10^{6}-1.58\times
10^{7}$ $M_{\odot }$kpc$^{-3}$. For a more comprehensive discussion on the
distribution of dark matter, see [136]. The local dark matter energy density,
consistent with standard estimates, is $\rho _{\odot }=(0.3\pm 0.1)$ GeV/cm$%
^{3}=7.92\times 10^{6}$ $M_{\odot }$kpc$^{-3}$ [137]. Bergstrom, Ullio and
Buckley [69] find local dark matter densities acceptable in a somewhat broad
range $0.2-0.8$ GeV/cm$^{3}$. The fitting with Burkert profile yields $\rho
_{\odot }^{\text{\textit{BP}}}=0.487$ GeV/cm$^{3}=1.28\times 10^{7}$ $%
M_{\odot }$kpc$^{-3}$ and fitting with NFW profile yields $\rho _{\odot }^{%
\text{\textit{NFW}}}=0.471$ GeV/cm$^{3}=1.24\times 10^{7}$ $M_{\odot }$kpc$%
^{-3}$ [132]. About systematic uncertainties in the determination of local
density of dark matter, see [139]. Overall, one could fairly say that $\rho
_{\odot }\varpropto 10^{6}-10^{7}$ $M_{\odot }$kpc$^{-3}$.

We use the above local values as a constraint to estimate the central
density $\rho _{0}^{\text{\textit{BEC}}}$ given in Eq.(40) that approaches a
constant value $\rho _{0}^{\text{\textit{BEC}}}$ as $r\rightarrow 0$ (so
does the EiBI profile $\rho ^{(0)}(r)$ in Eq.(3) as the two are essentially
of the same form). This behavior is consistent with the currently favored
core behavior at the galactic center, as opposed to the NFW cusp. To
estimate the values of $\rho _{0}^{\text{\textit{BEC}}}$ for the Milky Way,
we constrain the BEC profile such that it coincides with the local value $%
\rho _{\odot }^{\text{\textit{BP}}}=1.28\times 10^{7}$ $M_{\odot }$kpc$^{-3}$
at $R_{\odot }=8.5$ kpc (boundary condition). This then yields a central
density $\rho _{0}^{\text{\textit{BEC}}}=1.29\times 10^{7}$ $M_{\odot }$kpc$%
^{-3}$. This is quite an acceptable central density value for the Milky Way.
(We could as well use the same boundary condition using $\rho _{\odot }^{%
\text{\textit{NFW}}}$, but it does not lead to a much different value for $%
\rho _{0}^{\text{\textit{BEC}}}$). Having determined the value of $\rho
_{0}^{\text{\textit{BEC}}}$, we compare it with the corresponding values
from NFW and Burkert profiles using data from [132] and observe the
following: (i) $\rho _{0}^{\text{\textit{BEC}}}$ is quite comparable with $%
\rho _{0}^{\text{\textit{BP}}}$ and $\rho _{0}^{\text{\textit{NFW}}}$, (ii)
the NFW cusp and the PI, Burkert core behavior are evident from Fig. 6.
(iii) Identifying the BEC constant $\xi \equiv \frac{\pi }{R_{\text{DM}}}$,
we see that the $\rho ^{\text{\textit{BEC}}}$ profile shows a \textit{much
slower} monotonic decline from its central value $\rho _{0}^{\text{\textit{%
BEC}}}$, coasting along almost flat all the way up to a finite $R_{\text{DM}%
} $ ($=111.90$ kpc, in the present case), where it vanishes, (iv) adopted
values from the Burkert profile has allowed us to predict $\rho _{0}^{\text{%
\textit{BEC}}}$, which is seen also to be included in the limiting interval, 
$\rho _{0}^{\text{lower}}<\rho _{0}^{\text{\textit{BEC,BP,NFW}}}<\rho _{0}^{%
\text{upper}}$ for the Milky Way.

Returning to the profiles (39) and (40), it is remarkable that the PI and
BEC profiles have the same behavior up to second order in $r$ provided we
identify $r_{c}=$ $\frac{\sqrt{6}}{\xi }$ but they begin to differ in the
higher order coefficients thereafter. Also, it is known that the large
majority of the high-resolution rotation curves prefer the PI core-dominated
halo model, which provide a better description of the data than the cuspy ($%
\rho ^{\text{\textit{NFW}}}\propto r^{-1}$) NFW profile [140]. In this sense,
the EiBI model could be a competing candidate to PI model. It would be our
future task to investigate where these two models agree and where they
disagree.

\begin{center}
\textbf{VI. CONCLUSIONS}
\end{center}

The present paper, based on a pivotal input from Weyl gravity, viz., $R_{%
\text{DM}}=R_{\text{WR}}$ (motivated by [89]), offers a new alternative 
\textit{analytical} window, different from the standard data-fit approaches,
to look at the physical galactic parameters. While the latter approaches are
technically more elaborate, and the current EiBI analytic approach is not,
the value of the paper lies in the fact that it can still make quantitative
predictions about the limits on central dark matter density $\rho _{0}$.
Many samples for which the values of $\rho _{0}$ are available are shown to
satisfy the inequality $\rho _{0}\leq \rho _{0}^{\text{upper}}\propto R_{%
\text{WR}}^{-2}\sim 10^{12}$ $M_{\odot }$kpc$^{-3}$. Only some samples are
tabulated here. Table I shows the halo/dark matter radius, the velocity
terminating radius $R_{\text{term}}$ and the corresponding coupling
parameter $\kappa $.

Going a step further, we also calculated $\rho _{0}^{\text{lower}}$ $\propto
(\alpha -1)M_{\text{lum}}R_{\text{WR}}^{-3}$ that depends on a certain
parameter $\alpha $ equal to the the ratio of luminous to total (dark matter
included) mass of a galaxy. Definitive estimates of such ratios are yet
unavailable. Nevertheless, it is shown that (Table II), for reasonably wider
values of $\alpha $ ($\geq $ $12.5$) accounting for huge quantities of dark
matter in individual samples, profile dependent values of $\rho _{0}$ still
fall inside the EiBI predicted interval $\rho _{0}^{\text{lower}}\leq \rho
_{0}\leq \rho _{0}^{\text{upper}}$. These limits cover a large class of
galaxies and indicate an interesting facet of the EiBI theory. We especially
point out that the maximal value $\rho _{0}^{\text{upper}}\propto R_{\text{WR%
}}^{-2}\sim 10^{12}$ $M_{\odot }$kpc$^{-3}$ is purely a stability induced
constraint on all galaxies with dark matter, while $(\alpha -1)M_{\text{lum}%
}R_{\text{WR}}^{-3}$ $\propto \rho _{0}^{\text{lower}}\leq \rho _{0}$ is
not, due to uncertainties in $\alpha $. Thus, we would particularly advocate
a practical verification of $\rho _{0}^{\text{upper}}$ ($\propto 1/\kappa $)
rather than $\rho _{0}^{\text{lower}}$. If verified, it would also mean that
we have a clearcut theoretical algorithm, applicable to all galactic
samples, that provides a definitive, falsifiable information on the radius $%
R_{\text{DM}}$ of dark matter/halo $-$ something that seems rather scarce in
the astrophysical literature.

A special merit of the foregoing analyses is that the only information
needed to calculate the above limits are those of the fitted luminous $M_{%
\text{lum}}$ values and the measured total mass $M_{\text{tot}}$. Note that
a small change in $M_{\text{lum}}$ would lead to a large change in $R_{\text{%
WR}}$. For instance, there is an argument [117,141] for an upper mass limit
indicated by the sudden decline of the visible baryonic mass function of
disk galaxies at $M_{\text{disc}}^{\text{max}}=2\times 10^{11}M_{\odot }$.
Tentatively assuming that the luminous part of the Milky Way mass $M_{\text{%
lum}}$ is $2\times 10^{11}M_{\odot }$ instead of $1.04\times 10^{11}M_{\odot
}$, then the resultant $R_{\text{WR}}$ would jump to $177.94$ kpc from $%
111.90$ kpc. Similarly, $R_{\text{term }}$would jump to $238$ kpc from $%
150.17$ kpc. Thus, for reliable values of $R_{\text{WR}}$, the luminous mass
data $M_{\text{lum}}$ should be as accurate as possible.

We have verified that quantitative upper limit $\rho _{0}^{\text{upper}}\sim
10^{12}$ $M_{\odot }$kpc$^{-3}$ is respected by all the samples collected in
[107], some of which are given in Table II. The reason for such consistency
is not accidental $-$ it stemmed from the fact that the Weyl radius $R_{%
\text{WR}}$ has a solid foundation: The rotation curve is a \textit{%
prediction} of Weyl gravity MO solution containing constants ($\gamma _{0},$ 
$\gamma ^{\ast }$, $k$) that are universally applicable to all the galaxies%
\footnote{%
The claim is grounded to the fact that a single set of universal constants ($%
\gamma _{0},\gamma ^{\ast },k$) and the ($M/L$) ratio of individual samples,
all\textit{\ a priori }known, are enough to nicely predict all the rotation
data $-$ no adjustable free parameters are needed. In contradistinction,
NFW, Burkert, PI or other profiles only give the generic shapes of halos,
and leave the values for $\rho _{0}$ and $r_{0}$ as free parameters to be
finely tuned to data galaxy by galaxy. This procedure then quickly generates
large numbers of such values as more and more galactic rotation curves are
considered (For details of such "fine tuning", see [35], pp. 32,33).}, LSB or
HSB, and that the $R_{\text{WR}}$ is a straightforward result from $%
V^{\prime \prime }<0$. In fact, the reported data on $R_{\text{last}}$ for
individual samples have so far been found to obey $R_{\text{last}}<R_{\text{%
WR}}$. So we conjectured that this radius $R_{\text{WR}}$ just might be 
\textit{the} dark matter radius $R_{\text{DM}}$ specific to individual
galaxies.

As we saw, the constraint $\rho _{0}\leq \rho _{0}^{\text{upper}}\sim
10^{12} $ $M_{\odot }$kpc$^{-3}$ is a necessary condition for stability of
circular orbits. Whether it is also a sufficient condition, that is, whether
there are no stable orbits in the halo if this constraint is violated, is a
matter of independent practical verification. If sufficiency turns out to be
true, then we might expect to observe galaxies with no information on dark
matter due to lack of stable circular orbits. It may be noted that our $\rho
_{0}^{\text{upper}}\sim 6.14\times 10^{11}$ $M_{\odot }$kpc$^{-3}$ for Milky
Way is remarkably consistent with the\textit{\ local upper limit} on the
dark matter density in the solar system, $\rho _{\odot }^{\text{upper}}\sim
2.94\times 10^{12}$ $M_{\odot }$kpc$^{-3}$, found by completely different
methods and ideas [142].

There are limitations with almost all well known density profiles in the
sense that they fit the data so well in one sector, but fail in the other.
For instance, it has been argued [143] that the NFW profile does not always
follow from the gas rotation curves of large samples. For a constant
velocity anisotropy, the PI profile is ruled out, while a truncated flat
(TF) model [144] and NFW model are consistent with the data. Incidentally, it
might be noted that the TF model expands up to $r^{2}$ like both in the PI
and BEC profiles, and further, like the MO model, TF is described solely by
two parameters, mass and the scale length. Nesti and Salucci [132] argue that
NFW and/or PI halos are not supported by present day observations in
external galaxies due to recent improvement of simulation techniques. URC
profile for velocity distribution seems to fit the data incredibly well up
to $\sim 30$ kpc [97]. The present model based on EiBI Eq.(3), which is
akin to the quantum BEC model, is probably no better or worse than the
others. Nevertheless, the foregoing study hopefully provides some new
definitive information in an analytic way using a metric solution (7) of
EiBI theory.

\newpage

\begin{center}
\textbf{Table I. Lower bound on average density [eq. (36)}]. \\[0pt]
\bigskip 
\begin{tabular}{lllllll}
Galaxy & $R_{\text{WR}}$ & $R_{\text{term}}$ & $\kappa $ & $\left\langle
\rho \right\rangle ^{\text{lower}}$ & $\gamma $ & $\beta =\alpha \gamma $ \\ 
$\alpha =12.5$ & (kpc) & (kpc) & (kpc)$^{2}$ & $\text{(}M_{\odot }\text{kpc}%
^{-3}\text{)}$ &  &  \\ 
&  &  &  &  &  &  \\ 
ESO1870510 & $39.66$ & $52.96$ & $318.83$ & $2.45\times 10^{4}$ & $1.62$ & $%
20.25$ \\ 
&  &  &  &  &  &  \\ 
ESO3020120 & $44.77$ & $60.15$ & $406.11$ & $1.71\times 10^{4}$ & $1.07$ & $%
13.37$ \\ 
&  &  &  &  &  &  \\ 
ESO3050090 & $39.45$ & $52.65$ & $315.37$ & $2.49\times 10^{4}$ & $0.32$ & $%
04.00$ \\ 
&  &  &  &  &  &  \\ 
ESO4880490 & $42.25$ & $56.62$ & $361.70$ & $2.03\times 10^{4}$ & $3.07$ & $%
38.37$ \\ 
&  &  &  &  &  &  \\ 
U04115 & $39.02$ & $52.03$ & $308.53$ & $2.58\times 10^{4}$ & $0.97$ & $12.12
$ \\ 
&  &  &  &  &  &  \\ 
U11557 & $41.75$ & $55.91$ & $353.21$ & $2.11\times 10^{4}$ & $0.20$ & $02.50
$ \\ 
&  &  &  &  &  &  \\ 
U11748 & $106.18$ & $142.57$ & $2284.84$ & $1.28\times 10^{3}$ & $0.40$ & $%
05.00$ \\ 
&  &  &  &  &  &  \\ 
U11819 & $72.80$ & $98.08$ & $1073.95$ & $3.97\times 10^{3}$ & $2.24$ & $%
28.00$ \\ 
&  &  &  &  &  &  \\ 
U11583 & $39.08$ & $52.11$ & $309.47$ & $2.57\times 10^{4}$ & $0.70$ & $08.75
$ \\ 
&  &  &  &  &  &  \\ 
F568-3 & $50.10$ & $67.53$ & $508.76$ & $1.22\times 10^{4}$ & $4.2$ & $52.50$
\\ 
&  &  &  &  &  &  \\ 
F583-1 & $41.50$ & $55.57$ & $349.11$ & $2.14\times 10^{4}$ & $1.6$ & $20.00$%
\end{tabular}

\newpage \textbf{Table II. Central densities of dark matter and upper bounds [eq. (31)].
All densities are in units of $M_{\odot }$kpc$^{-3}$, $\alpha=12.5$}\\[0pt].
\bigskip

\begin{tabular}{llllll}

Galaxy & $\rho _{0}^{\text{lower}}$ & $\rho _{0}^{\text{BEC}}$ & $\rho _{0}^{\text{PI}}$ & $\rho _{0}^{\text{NFW}}$ & $\rho _{0}^{\text{upper}}$ \\ 
&  &  &  &  &  \\ 
ESO & $8.08\times 10^{4}$ & $3.29\times10^{7}$ & $5.48\times 10^{7}$ & $7.61\times 10^{5}$ & $4.88\times 10^{12}$ \\ 
1870510 &  &  &  &  &  \\ 
&  &  &  &  &  \\ 
ESO & $5.63\times 10^{4}$ & $2.29\times 10^{7}$ & $5.98\times 10^{7}$ & $2.65\times 10^{6}$ & $3.83\times 10^{12}$ \\ 
3020120 &  &  &  &  &  \\ 
&  &  &  &  &  \\ 
ESO & $8.22\times 10^{4}$ & $2.17\times 10^{7}$ & $2.76\times 10^{7}$ & $3.28\times 10^{7}$ & $4.93\times 10^{12}$ \\ 
3050090 &  &  &  &  &  \\ 
&  &  &  &  &  \\ 
ESO & $6.69\times 10^{4}$ & $5.49\times 10^{7}$ & $1.03\times 10^{8}$ & $1.42\times 10^{6}$ & $4.30\times 10^{12}$ \\ 
4880490 &  &  &  &  &  \\ 
&  &  &  &  &  \\ 
U04115 & $8.49\times 10^{4}$ & $1.43\times 10^{8}$ & $1.51\times 10^{8}$ & $1.39\times 10^{5}$ & $5.04\times 10^{12}$ \\ 
&  &  &  &  &  \\ 
U11557 & $6.94\times 10^{4}$ & $4.69\times 10^{9}$ & $1.57\times 10^{7}$ & $1.08\times 10^{4}$ & $4.41\times 10^{12}$ \\ 
&  &  &  &  &  \\ 
U11748 & $4.21\times 10^{3}$ & $4.20\times 10^{8}$ & $1.67\times 10^{9}$ & $2.04\times 10^{8}$ & $6.81\times 10^{11}$ \\ 
&  &  &  &  &  \\ 
U11819 & $1.30\times 10^{3}$ & $5.39\times 10^{7}$ & $8.69\times 10^{7}$ & $1.19\times 10^{6}$ & $1.45\times 10^{12}$ \\ 
&  &  &  &  &  \\ 
U11583 & $8.45\times 10^{4}$ & $9.53\times 10^{7}$ & $1.19\times 10^{8}$ & $1.36\times 10^{5}$ & $5.03\times 10^{12}$ \\ 
&  &  &  &  &  \\ 
F568-3 & $4.01\times 10^{4}$ & $2.48\times 10^{7}$ & $3.61\times 10^{7}$ & $3.78\times 10^{5}$ & $3.06\times 10^{12}$ \\ 
&  &  &  &  &  \\ 
F583-1 & $7.05\times 10^{4}$ & $1.90\times 10^{7}$ & $3.17\times 10^{7}$ & $3.45\times 10^{5}$ & $4.46\times 10^{12}$%
\end{tabular}
\end{center}

\newpage

\begin{center}
\textbf{Figure captions}
\end{center}

\begin{figure}[h]
\includegraphics [width=\linewidth] {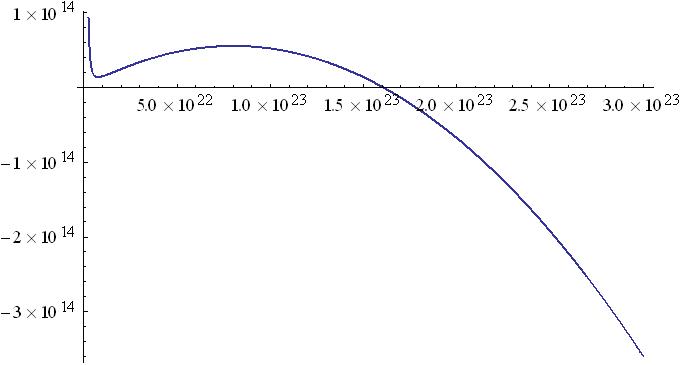}
\caption{Plot of $v_{\text{tg}}^{2}$ vs $R$ (cm) using Eq.(15) for
ESO 1200211. The luminous mass is $M_{\text{lum}}=5.6\times 10^{7}M_{\odot }$
so that $N^{\ast }=5.6\times 10^{7}$. The other constants are in (16) and $%
R_{0}=2$ kpc [44]. One finds the velocity terminating radius $R_{\text{term}%
}=52.04$ kpc.}
\end{figure}
\bigskip 
\begin{figure}[tbp]
\includegraphics [width=\linewidth] {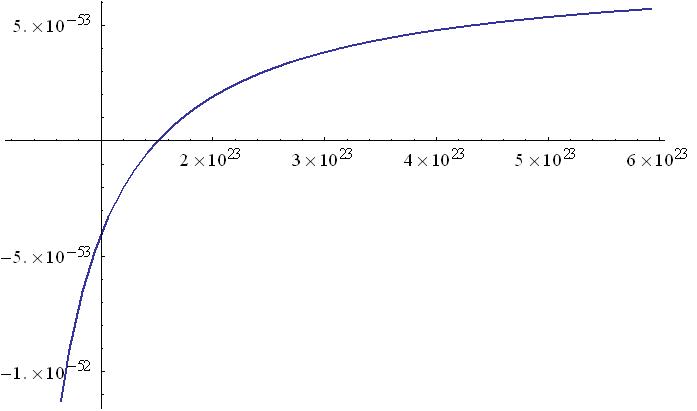}
\caption{Plot of $V^{\prime \prime }$ vs $R$ (cm) using Eq.(21). The
crossing shows the halo radius $R_{\text{WR}}=39.033$ kpc ($\equiv
1.204\times 10^{23} $ cm), which is the maximally allowed radius supporting
stable circular orbits in the halo of ESO 1200211, for which $R_{0}=2$ kpc.
The plot is made for the radii $R>4R_{0}.$}
\end{figure}
\bigskip 
\begin{figure}[tbp]
\includegraphics [width=\linewidth] {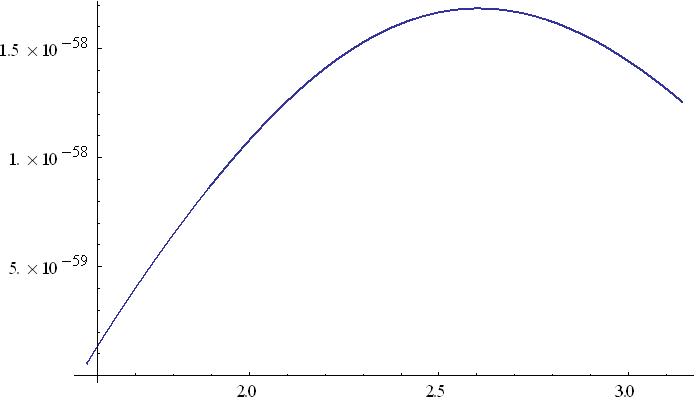}
\caption{Plot of $V^{\prime \prime }$ vs $\overline{R}\in \lbrack 0.5\protect%
\pi ,\protect\pi ] $ using Eq.(30) for ESO 1200211. The chosen parameters
are: $r_{0}=0.61$, $v_{\infty }^{2}=0.000001$, $r_{\text{opt}}=8$ kpc ($%
\equiv 2.47\times 10^{22} $ cm). Here $\protect\rho _{0}=8.25$ $\times
10^{12}$ $M_{\odot }$kpc$^{-3}$, which corresponds to $\overline{\protect%
\rho }_{0}=0.50\protect\pi $. The orbits are unstable in the chosen entire
radial range $\overline{R}$ $\in \lbrack 0.5\protect\pi ,\protect\pi ]$
because $V^{\prime \prime }>0$ there.}
\end{figure}
\bigskip 
\begin{figure}[tbp]
\includegraphics [width=\linewidth] {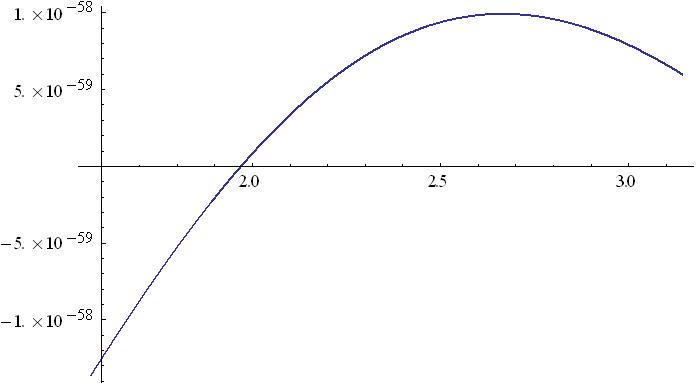}
\caption{Plot of $V^{\prime \prime }$ vs $\overline{R}\in \lbrack 0.5\protect%
\pi ,\protect\pi ] $ using Eq.(30) for ESO 1200211. The chosen parameters
are: $r_{0}=0.61$, $v_{\infty }^{2}=0.000001$, $r_{\text{opt}}=8$ kpc. Here
central density is further lowered to $\protect\rho _{0}=5.61\times 10^{12}$ 
$M_{\odot }$kpc$^{-3}$, which corresponds to $\overline{\protect\rho }%
_{0}=0.34\protect\pi $. The orbit are unstable in some intermediate radii as 
$V^{\prime \prime }$ is partly positive and partly negative.}
\end{figure}
\bigskip 
\begin{figure}[tbp]
\includegraphics [width=\linewidth] {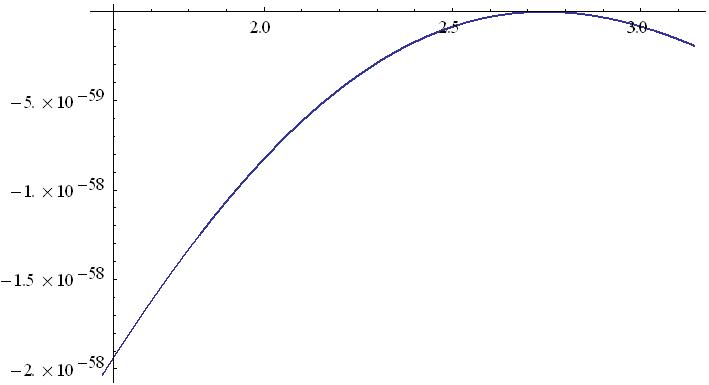}
\caption{Plot of $V^{\prime \prime }$ vs $\overline{R}\in \lbrack 0.5\protect%
\pi ,\protect\pi ] $ using Eq.(30) for ESO 1200211. The chosen parameters
are: $r_{0}=0.61$, $v_{\infty }^{2}=0.000001$, $r_{\text{opt}}=8$ kpc. Here
central density is further lowered to $\protect\rho _{0}=5.04$ $\times
10^{12}$ $M_{\odot }$kpc$^{-3}$, which corresponds to $\overline{\protect%
\rho }_{0}^{\text{upper}}=$ $0.94=\protect\beta ^{\text{upper}}\protect\pi $%
, where $\protect\beta ^{\text{upper}}=0.299$. The orbit is \textit{stable}
in the entire chosen range for $\overline{R}$. The corresponding $\protect%
\rho _{0}$ is the upper limit on central density $\protect\rho _{0}^{\text{%
upper}}$ specific to the sample.}
\end{figure}
\bigskip 
\begin{figure}[tbp]
\includegraphics [width=\linewidth] {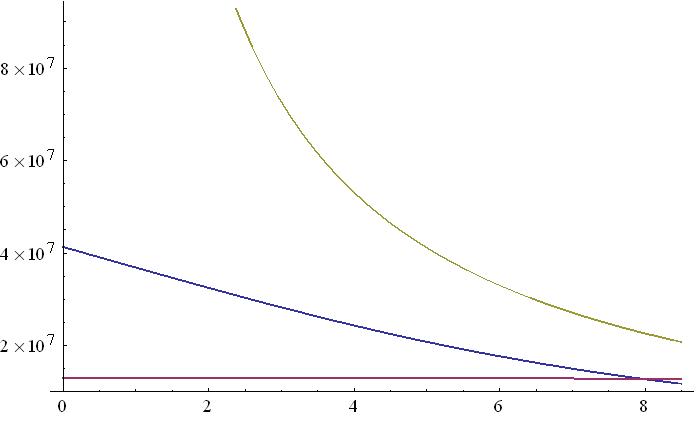}
\caption{$\rho (r)$ \textit{vs} $r$ for three profiles for the Milky Way.
Bottom brown is BEC profile ($\rho _{0}^{\text{\textit{BEC}}}=1.292\times
10^{7}$ $M_{\odot }$kpc$^{-3}$, $\xi =\frac{\pi }{R_{\text{DM}}}=0.028$ kpc$%
^{-1}$), Blue curve is Burkert profile ($\rho _{0}^{\text{\textit{BP}}%
}=4.13\times 10^{7}$ $M_{\odot }$kpc$^{-3}$, $r_{0}=9.26$ kpc) and the green
one is NFW profile ($\rho _{0}^{\text{\textit{NFW}}}=1.40\times 10^{7}$ $%
M_{\odot }$kpc$^{-3}$, $r_{S}=16.1$ kpc). Uncertainties shown in the
original fitted values not considered. Data taken from [132].}
\end{figure}

\newpage

\bigskip

\textbf{Acknowledgments}

One of us (Ramil Izmailov) was supported by the Ministry of Education and
Science of Russian Federation. This work was supported in part by an
internal grant of M. Akmullah Bashkir State Pedagogical University in the
field of natural sciences. The authors are thankful to Guzel Kutdusova,
Regina Lukmanova and Almir Yanbekov for technical assistance.

\bigskip

\textbf{REFERENCES}

[1] J. Oort, Some Problems Concerning the Distribution of Luminosities and Peculiar Velocities of Extragalactic Nebulae, Bull. Astron. Inst. Netherlands 6 (1931) 155.

[2] F. Zwicky, Die Rotverschiebung von extragalaktischen Nebeln, Helv. Phys. Acta 6 (1933) 110.

[3] F. Zwicky, On the Masses of Nebulae and of Clusters of Nebulae, Astrophys. J. 86 (1937) 217.

[4] K.C. Freeman, On the Disks of Spiral and so Galaxies, Astrophys. J. 160 (1970) 881.

[5] M.S. Roberts and A.H. Rots, Comparison of Rotation Curves of Different Galaxy Types, Astron. Astrophys. 26 (1973) 483.

[6] P. Ostriker, P.J.E. Peebles and A. Yahill, The size and mass of galaxies, and the mass of the universe, Astrophys. J. Lett. 193 (1974) L1.

[7] J. Einasto, A. Kaasik and E. Saar, Dynamic evidence on massive coronas of galaxies, Nature 250 (1974) 309.

[8] V.C. Rubin, N. Thonnard and W.K. Ford Jr., Extended rotation curves of high-luminosity spiral galaxies. IV. Systematic dynamical properties, Sa through Sc, Astrophys. J. Lett. 225
(1978) L107.

[9] V.C. Rubin, M.S. Roberts and W.K. Ford Jr., Extended rotation curves of high-luminosity spiral galaxies. V. NGC 1961, the most massive spiral known, Astrophys. J. 230 (1979) 35.

[10] Y. Sofue and V. Rubin, Rotation curves of spiral galaxies, Ann. Rev. Astron. Astrophys. 39 (2001) 137 [astro-ph/0010594].

[11] E. Maoz, Gravitational microlensing by dark clusters in the galactic halo, Astrophys. J. 428 (1994) L5.

[12] D.G. Barnes, R.L. Webster, R.W. Schmidt and A. Hughes, Imaging H I in the lensing galaxy 2237 + 0305, Mon. Not. Roy. Astron. Soc. 309 (1999) 641.

[13] Y.-C.N. Cheng and L.M. Krauss, Gravitational lensing and dark structures, Astrophys. J. 514 (1999) 25.

[14] C.M. Trott and R.L. Webster, Dissecting a galaxy: mass distribution of 2237 + 0305, Mon. Not. Roy. Astron. Soc. 334 (2002) 621.

[15] N.N. Weinberg and M. Kamionkowski, Weak gravitational lensing by dark clusters, Mon. Not. Roy. Astron. Soc. 337 (2002) 1269.

[16] R.J. Smith, J.P. Blakeslee, J.R. Lucey and J. Tonry, Discovery of strong lensing by an elliptical galaxy at z = 0.0345, Astrophys. J. 625 (2005) L103.

[17] T. Faber and M. Visser, Combining rotation curves and gravitational lensing: How to measure the equation ofstate ofdark matter in the galactic halo,
Mon. Not. Roy. Astron. Soc. 372 (2006) 136.

[18] R.B. Metcalf and J. Silk, New Constraints on Macroscopic Compact Objects as a Dark Matter Candidate from Gravitational Lensing ofType Ia Supernovae,
Phys. Rev. Lett. 98 (2007) 071302.

[19] S. Bharadwaj and S. Kar, Modeling galaxy halos using dark matter with pressure, Phys. Rev. D 68 (2003) 023516.

[20] M. Colpi, S.L. Shapiro and I. Wasserman, Boson Stars: Gravitational Equilibria of Selfinteracting Scalar Fields, Phys. Rev. Lett. 57 (1986) 2485.

[21] G. Efstathiou, W.J. Sutherland and S.J. Maddox, The cosmological constant and cold dark matter, Nature 348 (1990) 705.

[22] SDSS collaboration, A.C. Pope et al., Cosmological parameters from Eigenmode analysis of Sloan Digital Sky Survey galaxy redshifts, Astrophys. J. 607 (2004) 655.

[23] SDSS collaboration, M. Tegmark et al., Cosmological parameters from SDSS and WMAP, Phys. Rev. D 69 (2004) 103501.

[24] SDSS collaboration, M. Tegmark et al., The three-dimensional power spectrum of galaxies from the SDSS, Astrophys. J. 606 (2004) 702.

[25] T. Matos, F.S. Guzman and D. Nunez, Spherical scalar field halo in galaxies, Phys. Rev. D 62 (2000) 061301.

[26] P.J.E. Peebles, Dynamics of a dark matter field with a quartic selfinteraction potential, Phys. Rev. D 62 (2000) 023502.

[27] T. Matos and F.S. Guzman, Scalar fields as dark matter in spiral galaxies, Class. Quant. Grav. 17 (2000) L9 [gr-qc/9810028].

[28] U. Nucamendi, M. Salgado and D. Sudarsky, An Alternative approach to the galactic dark matter problem, Phys. Rev. D 63 (2001) 125016 [gr-qc/0011049].

[29] E.W. Mielke and F.E. Schunck, Nontopological scalar soliton as dark matter halo, Phys. Rev. D 66 (2002) 023503.

[30] L.G. Cabral-Rosetti, T. Matos, D. Nunez and R.A. Sussman, Hydrodynamics of galactic dark matter, Class. Quant. Grav. 19 (2002) 3603 [gr-qc/0112044].

[31] J.E. Lidsey, T. Matos and L.A. Urena-Lopez, The Inflaton field as selfinteracting dark matter in the brane world scenario, Phys. Rev. D 66 (2002) 023514.

[32] A. Arbey, J. Lesgourgues and P. Salati, Galactic halos of fluid dark matter, Phys. Rev. D 68 (2003) 023511.

[33] M.K. Mak and T. Harko, Can the galactic rotation curves be explained in brane world models?, Phys. Rev. D 70 (2004) 024010 [gr-qc/0404104].

[34] A. Borriello and P. Salucci, The Dark matter distribution in disk galaxies, Mon. Not. Roy. Astron. Soc. 323 (2001) 285.

[35] P.D. Mannheim, Alternatives to dark matter and dark energy, Prog. Pari. Nucl. Phys. 56 (2006) 340.

[36] M. Milgrom, A Modification of the Newtonian dynamics as a possible alternative to the hidden mass hypothesis, Astrophys. J. 270 (1983) 365.

[37] M. Milgrom, A Modification of the Newtonian dynamics: Implications for galaxies, Astrophys. J. 270 (1983) 371.

[38] M. Milgrom, A modification of the Newtonian dynamics: implications for galaxy systems, Astrophys. J. 270 (1983) 384.

[39] R.H. Sanders, The published extended rotation curves of spiral galaxies: confrontation with modified dynamics, Astrophys. J. 473 (1996) 117.

[40] R.A. Swaters, R.H. Sanders and S.S. McGaugh, Testing Modified Newtonian Dynamics with Rotation Curves of Dwarfand Low Surface Brightness Galaxies, Astrophys. J. 718 (2010) 380 [arXiv:1005.5456].

[41] G. Gentile, B. Famaey, F. Combes, P. Kroupa, H.S. Zhao nad O. Tiret, Tidal dwarf galaxies as a test of fundamental physics, Astron. Astrophys. 472 (2007) L25 [arXiv:0706.1976].

[42] L. Iorio, Galactic Sun's motion in the cold dark matter, MOdified Newtonian Dynamics and modified gravity scenarios, Astron. Nachr. 330 (2009) 857.

[43] J.W. Moffat, Scalar-tensor-vector gravity theory, JCAP 03 (2006) 004 [gr-qc/0506021].

[44] J.R. Brownstein and J.W. Moffat, Galaxy rotation curves without non-baryonic dark matter, Astrophys. J. 636 (2006) 721.

[45] S. Capozziello, V.F. Cardone and A. Troisi, Low surface brightness galaxies rotation curves in the low energy limit of Rn gravity: no need for dark matter?,
Mon. Not. Roy. Astron. Soc. 375 (2007) 1423.

[46] E.E. Flanagan, Fourth order Weyl gravity, Phys. Rev. D 74 (2006) 023002.

[47] P.D. Mannheim, Schwarzschild limit ofconformal gravity in the presence of macroscopic scalar fields, Phys. Rev. D 75 (2007) 124006 [gr-qc/0703037].

[48] P.D. Mannheim and J.G. O'Brien, Impact of a global quadratic potential on galactic rotation curves, Phys. Rev. Lett. 106 (2011) 121101 [arXiv:1007.0970].

[49] G.R. Blumenthal, S.M. Faber, J.R. Primack and M.J. Rees, Formation of Galaxies and Large Scale Structure with Cold Dark Matter, Nature 311 (1984) 517.

[50] A.R. Liddle and D.H. Lyth, The Cold dark matter density perturbation, Phys. Rept. 231 (1993).

[51] P.J.E. Peebles, Large scale background temperature and mass fluctuations due to scale invariant primeval perturbations, Astrophys. J. 263 (1982) L1.

[52] P.J.E. Peebles, Dark matter and the origin of galaxies and globular star clusters, Astrophys. J. 277 (1984) 470.

[53] S. Basilakos and J.A.S. Lima, Constraints on Cold Dark Matter Accelerating Cosmologies and Cluster Formation, Phys. Rev. D 82 (2010) 023504 [arXiv:1003.5754].

[54] S. McGaugh, Dynamics and the second peak: Cold dark matter?, Int. J. Mod. Phys. A 16S1C (2001) 1031.

[55] C.S. Kochanek, E.E. Falco, C. Impey, J. Lehar, B. McLeod and H.-W. Rix, CASTLE Survey Gravitational Lens Data Base (2005).

[56] V.K. Onemli, Probing Cold Dark Matter Cusps by Gravitational Lensing, Int. J. Mod. Phys. D 15 (2006) 2059.

[57] S. Dodelson, E.I. Gates and M.S. Turner, Cold dark matter models, Science 274 (1996) 69.

[58] J.-c. Hwang and H. Noh, Axion as a Cold Dark Matter candidate, Phys. Lett. B 680 (2009) 1 [arXiv:0902.4738].

[59] S.L. Dubovsky, P.G. Tinyakov and I.I. Tkachev, Massive graviton as a testable cold dark matter candidate, Phys. Rev. Lett. 94 (2005) 181102.

[60] C. Evoli, A. Mesinger and A. Ferrara, Unveiling the nature of dark matter with high redshift 21 cm line experiments, JCAP 11 (2014) 024 [arXiv:1408.1109].

[61] O.F. Piattella, D.L.A. Martins and L. Casarini, Sub-horizon evolution of cold dark matter perturbations through dark matter-dark energy equivalence epoch, JCAP 10 (2014) 031 [arXiv:1407.4773].

[62] C.M. Ho, D. Minic and Y.J. Ng, Cold Dark Matter with MOND Scaling, Phys. Lett. B 693 (2010) 567 [arXiv:1005.3537].

[63] C. Rampf and G. Rigopoulos, Initial conditions for cold dark matter particles and General Relativity, Phys. Rev. D 87 (2013) 123525 [arXiv:1305.0010].

[64] A.A. Grib and Y. Pavlov, Cold dark matter and primordial superheavy particles, Int. J. Mod. Phys. A 17 (2002) 4435 [gr-qc/0211015].

[65] A. Zhitnitsky, Cold dark matter as compact composite objects, Phys. Rev. D 74 (2006) 043515.

[66] H. Li, J. Liu, J.-Q. Xia and Y.-F. Cai, Cold Dark Matter Isocurvature Perturbations: Cosmological Constraints and Applications, Phys. Rev. D 83 (2011) 123517 [arXiv:1012.2511].

[67] AMS collaboration, M. Aguilar et al., First Result from the Alpha Magnetic Spectrometer on the International Space Station: Precision Measurement of the Positron Fraction in Primary Cosmic Rays of 0.5-350 GeV, Phys. Rev. Lett. 110 (2013) 141102.

[68] A. De Simone, A. Riotto and W. Xue, Interpretation of AMS-02 Results: Correlations among Dark Matter Signals, JCAP 05 (2013) 003 [arXiv:1304.1336].

[69] J. Binney and S. Tremaine, Galactic dynamics, Princeton University Press, Princeton U.S.A.(1987).

[70] M. Persic, P. Salucci and F. Stel, The Universal rotation curve of spiral galaxies: 1. The Dark matter connection, Mon. Not. Roy. Astron. Soc. 281 (1996) 27.

[71] H.J. de Vega, P. Salucci and N.G. Sanchez, The mass of the dark matter particle from theory and observations, New Astron. 17 (2012) 653 [arXiv:1004.1908].

[72] S. Nojiri and S.D. Odintsov, Gravity assisted dark energy dominance and cosmic acceleration, Phys. Lett. B 599 (2004) 137.

[73] G. Allemandi, A. Borowiec, M. Francaviglia and S.D. Odintsov, Dark energy dominance and cosmic acceleration in first order formalism, Phys. Rev. D 72 (2005) 063505 [gr-qc/0504057].

[74] S. Nojiri and S.D. Odintsov, Unified cosmic history in modified gravity: from F(R) theory to Lorentz non-invariant models, Phys. Rept. 505 (2011) 59 [arXiv:1011.0544].

[75] O. Bertolami, C.G. Boehmer, T. Harko and F.S.N. Lobo, Extra force in f (R) modified theories of gravity, Phys. Rev. D 75 (2007) 104016 [arXiv:0704.1733].

[76] T. Harko and F.S.N. Lobo, f (R, Lm) gravity, Eur. Phys. J. C 70 (2010) 373 [arXiv:1008.4193].

[77] T. Delsate and J. Steinhoff, New insights on the matter-gravity coupling paradigm, Phys. Rev. Lett. 109 (2012) 021101 [arXiv:1201.4989].

[78] M. Baiiados and P.G. Ferreira, Eddington's theory of gravity and its progeny, Phys. Rev. Lett. 105 (2010) 011101 [arXiv:1006.1769].

[79] S. Deser and G.W. Gibbons, Born-Infeld-Einstein actions?, Class. Quant. Grav. 15 (1998) L35.

[80] J. Casanellas, P. Pani, I. Lopes and V. Cardoso, Testing alternative theories of gravity using the Sun, Astrophys. J. 745 (2012) 15 [arXiv:1109.0249].

[81] P. Pani, V. Cardoso and T. Delsate, Compact stars in Eddington inspired gravity, Phys. Rev. Lett. 107 (2011) 031101 [arXiv:1106.3569].

[82] T. Harko, F.S.N. Lobo, M.K. Mak and S.V. Sushkov, Structure of neutron, quark and exotic stars in Eddington-inspired Born-Infeld gravity, Phys. Rev. D 88 (2013) 044032 [arXiv:1305.6770].

[83] J.H.C. Scargill, M. Banados and P.G. Ferreira, Cosmology with Eddington-inspired Gravity, Phys. Rev. D 86 (2012) 103533 [arXiv:1210.1521].

[84] P.P. Avelino and R.Z. Ferreira, Bouncing Eddington-inspired Born-Infeld cosmologies: an alternative to Inflation?, Phys. Rev. D 86 (2012) 041501 [arXiv:1205.6676].

[85] X.-L. Du, K. Yang, X.-H. Meng and Y.-X. Liu, Large Scale Structure Formation in Eddington-inspired Born-Infeld Gravity, Phys. Rev. D 90 (2014) 044054 [arXiv:1403.0083].

[86] Y.-X. Liu, K. Yang, H. Guo and Y. Zhong, Domain Wall Brane in Eddington Inspired Born-Infeld Gravity, Phys. Rev. D 85 (2012) 124053 [arXiv:1203.2349].

[87] Q.-M. Fu, L. Zhao, K. Yang, B.-M. Gu and Y.-X. Liu, Stability and (quasi)localization of gravitational fluctuations in an Eddington-inspired Born-Infeld brane system, Phys. Rev. D 90 (2014) 104007 [arXiv:1407.6107].

[88] S.-W. Wei, K. Yang and Y.-X. Liu, Black hole solution and strong gravitational lensing in Eddington-inspired Born-Infeld gravity, Eur. Phys. J. C 75 (2015) 253 [arXiv:1405.2178].

[89] A. Edery and M.B. Paranjape, Classical tests for Weyl gravity: Deflection of light and radar echo delay, Phys. Rev. D 58 (1998) 024011.

[90] A. Bhattacharya, A. Panchenko, M. Scalia, C. Cattani and K.K. Nandi, Light bending in the galactic halo by Rindler-Ishak method, JCAP 09 (2010) 004 [arXiv:0910.1112].

[91] A. Bhattacharya, G.M. Garipova, E. Laserra, A. Bhadra and K.K. Nandi, The Vacuole Model: New Terms in the Second Order Deflection of Light, JCAP 02 (2011) 028 [arXiv:1002.2601].

[92] D. Cutajar and K.Z. Adami, Strong lensing as a test for Conformal Weyl Gravity, Mon. Not. Roy. Astron. Soc. 441 (2014) 1291 [arXiv:1403.7930].

[93] F. Rahaman, K.K. Nandi, A. Bhadra, M. Kalam and K. Chakraborty, Perfect Fluid Dark Matter, Phys. Lett. B 694 (2010) 10 [arXiv:1009.3572].

[94] T. Harko and F.S.N. Lobo, Two-fluid dark matter models, Phys. Rev. D 83 (2011) 124051 [arXiv:1106.2642].

[95] T. Harko, F.S.N. Lobo, M.K. Mak and S.V. Sushkov, Dark matter density profile and galactic metric in Eddington-inspired Born-Infeld gravity, Mod. Phys. Lett. A 29 (2014) 1450049.

[96] P. Pani, T. Delsate and V. Cardoso, Eddington-inspired Born-Infeld gravity. Phenomenology of non-linear gravity-matter coupling, Phys. Rev. D 85 (2012) 084020 [arXiv:1201.2814].

[97] P. Salucci and A. Burkert, Dark matter scaling relations, Astrophys. J. 537 (2000) L9.

[98] P. Salucci and M. Persic, Maximal halos in high-luminosity spiral galaxies, Astron. Astrophys.351 (1999) 442.

[99] K.K. Nandi, A.I. Filippov, F. Rahaman, S. Ray, A.A. Usmani et al., Features of galactic halo in a brane world model and observational constraints,
Mon. Not. Roy. Astron. Soc. 399 (2009) 2079 [arXiv:0812.2545].

[100] K.K. Nandi and A. Bhadra, Comment on 'Impact of a Global Quadratic Potential on Galactic Rotation Curves', Phys. Rev. Lett. 109 (2012) 079001 [arXiv:1208.5330].

[101] K. Lake, Galactic potentials, Phys. Rev. Lett. 92 (2004) 051101 [gr-qc/0302067].

[102] D.N. Vollick, Palatini approach to Born-Infeld-Einstein theory and a geometric description of electrodynamics, Phys. Rev. D 69 (2004) 064030 [gr-qc/0309101].

[103] D.N. Vollick, Born-Infeld-Einstein theory with matter, Phys. Rev. D 72 (2005) 084026 [gr-qc/0506091]. 

[104] W.J.G. de Blok, S.S. McGaugh, A. Bosma and V.C. Rubin, Mass density profiles of LSB galaxies, Astrophys. J. 552 (2001) L23.

[105] S. Chandrasekhar, Mathematical Theory of Black Holes, Oxford University Press, Oxford U.K. (1983).

[106] P.D. Mannheim and D. Kazanas, Exact Vacuum Solution to Conformal Weyl Gravity and Galactic Rotation Curves, Astrophys. J. 342 (1989) 635.

[107] P.D. Mannheim and J.G. O'Brien, Fitting galactic rotation curves with conformal gravity and a global quadratic potential, Phys. Rev. D 85 (2012) 124020 [arXiv:1011.3495].

[108] V.H. Robles and T. Matos, Flat Central Density Profile and Constant DM Surface Density in Galaxies from Scalar Field Dark Matter, Mon. Not. Roy. Astron. Soc. 422 (2012) 282
[arXiv:1201.3032].

[109] R.H. Sanders, The published extended rotation curves of spiral galaxies: confrontation with modified dynamics, Astrophys. J. 473 (1996) 117 [astro-ph/9606089].

[110] D. Lynden-Bell and G.Gilmore eds., Baryonic Dark Matter, Kluwer, Dordrecht Netherlands (1990).

[111] K. Kuijken and G. Gilmore, The mass distribution in the galactic disc — I. A technique to determine the integral surface mass density ofthe disc near the Sun, Mon. Not. Roy. Astron.
Soc. 239 (1989) 571.

[112] K. Kuijken and G. Gilmore, The Mass Distribution in the Galactic Disc — II. Determination ofthe Surface Mass Density ofthe Galactic Disc Near the Sun, Mon. Not. Roy. Astron. Soc.
239 (1989) 605 .

[113] K. Kuijken and G. Gilmore, The mass distribution in the galactic disc — III. The local volume mass density, Mon. Not. Roy. Astron. Soc. 239 (1989) 651.

[114] K. Kuijken and G. Gilmore, The galactic disk surface mass density and the Galactic force K(z) at Z = 1.1 kiloparsecs, Astrophys. J. 367 (1991) L9.

[115] W.J.G. de Blok and S.S. McGaugh, The Dark and visible matter content of low surface brightness disk galaxies, Mon. Not. Roy. Astron. Soc. 290 (1997) 533.

[116] A. Burkert, The structure of dark matter halos in dwarf galaxies, Astrophys. J. 447 (1995) L25.

[117] P. Salucci, A. Lapi, C. Tonini, G. Gentile, I. Yegorova and U. Klein, The Universal Rotation Curve of Spiral Galaxies — II. The Dark Matter Distribution out to the Virial Radius, Mon. Not. Roy. Astron. Soc. 378 (2007) 41.

[118] S. Grossman and R. Narayan, Gravitationally lensed images in Abell 370, Astrophys. J. 344(1989) 637.

[119] M. Bartelmann and R. Narayan, Gravitational lensing and the mass distribution of clusters, AIP Conf. Proc. 336 (1995) 307.

[120] K.G. Begeman, A.H. Broeils and R.H. Sanders, Extended rotation curves of spiral galaxies: Dark haloes and modified dynamics, Mon. Not. Roy. Astron. Soc. 249 (1991) 523.

[121] J.F. Navarro, C.S. Frenk and S.D.M. White, The Structure of cold dark matter halos, Astrophys. J. 462 (1996) 563.

[122] J.F. Navarro, C.S. Frenk and S.D.M. White, A Universal density profile from hierarchical clustering, Astrophys. J. 490 (1997) 493.

[123] P.R. Kafle, S. Sharma, G.F. Lewis and J. Bland-Hawthorn, On the Shoulders of Giants: Properties of the Stellar Halo and the Milky Way Mass Distribution, Astrophys. J. 794 (2014) 59 [arXiv:1408.1787].

[124] S.M. Faber and J.S. Gallagher, Masses and mass-to-light ratios of galaxies, Ann. Rev. Astron. Astrophys. 17 (1979) 135.

[125] P.J. McMillan, Mass models of the Milky Way, Mon. Not. Roy. Astron. Soc. 414 (2011) 2446 [arXiv:1102.4340].

[126] F. Iocco, M. Pato, G. Bertone and P. Jetzer, Dark Matter distribution in the Milky Way:microlensing and dynamical constraints, JCAP 11 (2011) 029 [arXiv:1107.5810].

[127] A.J. Deason, V. Belokurov, N.W. Evans and J.H. An, Broken Degeneracies: The Rotation Curve and Velocity Anisotropy of the Milky Way Halo,
Mon. Not. Roy. Astron. Soc. 424 (2012) L44 [arXiv:1204.5189].

[128] A.J. Deason, V. Belokurov, N.W. Evans, S.E. Koposov, R.J. Cooke et al., The cold veil of the Milky Way stellar halo, Mon. Not. Roy. Astron. Soc. 425 (2012) 2840 [arXiv:1205.6203].
[129] L.E. Strigari, Galactic Searches for Dark Matter, Phys. Rept. 531 (2013) 1 [arXiv:1211.7090].

[130] G. Bertone, D. Hooper and J. Silk, Particle dark matter: Evidence, candidates and constraints, Phys. Rept. 405 (2005) 279.

[131] J.I. Read, The Local Dark Matter Density, J. Phys. G 41 (2014) 063101 [arXiv:1404.1938].

[132] F. Nesti and P. Salucci, The Dark Matter halo of the Milky Way, AD 2013, JCAP 07 (2013) 016 [arXiv:1304.5127].

[133] J.N. Bahcall, M. Schmidt and R.M. Soneira, The Galactic Spheroid, Astrophys. J. 265 (1983) 730.

[134] J.A.R. Caldwell and J.P. Ostriker, The Mass distribution within our Galaxy: A Three component model, Astrophys. J. 251 (1981) 61.

[135] M.S. Turner, Cosmic and Local Mass Density of Invisible Axions, Phys. Rev. D 33 (1986) 889.

[136] G. Jungman, M. Kamionkowski and K. Griest, Supersymmetric dark matter, Phys. Rept. 267 (1996) 195.

[137] J. Bovy and S. Tremaine, On the local dark matter density, Astrophys. J. 756 (2012) 89 [arXiv:1205.4033].

[138] L. Bergstrom, P. Ullio and J.H. Buckley, Observability of gamma-rays from dark matter neutralino annihilations in the Milky Way halo, Astropart. Phys. 9 (1998) 137.

[139] M. Pato, O. Agertz, G. Bertone, B. Moore and R. Teyssier, Systematic uncertainties in the determination of the local dark matter density, Phys. Rev. D 82 (2010) 023531 [arXiv:1006.1322].

[140] W.J.G. de Blok, S.S. McGaugh and V.C. Rubin, High-Resolution Rotation Curves ofLow Surface Brightness Galaxies. II. Mass Models, Astron. J. 122 (2001) 2396.

[141] P. Salucci and M. Persic, The Baryonic mass function of spiral galaxies: Clues to galaxy formation and to the nature ofdamped Lyman alpha clouds, Mon. Not. Roy. Astron. Soc. 309 (1999) 923.

[142] L. Iorio, Effect of Sun and Planet-Bound Dark Matter on Planet and Satellite Dynamics in the Solar System, JCAP 05 (2010) 018 [arXiv:1001.1697].
 
[143] G. Battaglia, A. Helmi, H. Morrison, P. Harding, E.W. Olszewski et al., The Radial velocity dispersion profile of the Galactic Halo: Constraining the density profile of the dark halo of the Milky Way, Mon. Not. Roy. Astron. Soc. 364 (2005) 433.

[144] M.I. Wilkinson and N.W. Evans, The present and future mass of the Milky Way halo, Mon. Not. Roy. Astron. Soc. 310 (1999) 645.

\bigskip

\bigskip

\bigskip

\end{document}